\documentclass{aa}  

\usepackage{graphicx, natbib, enumerate, txfonts}
\usepackage{color}

\usepackage{xstring}
\usepackage{catchfile}
\newcommand{\gitfilename}{./.git/refs/heads/master.}
\IfFileExists{\gitfilename}{%
  \newcommand{\gitrevision}{\StrLeft{\input{\gitfilename}}{7}}%
}

\def\table{(as used in Table \ref{tab:table})}

\def\GroupSolid{group \textit{solid}}
\def\GroupFluffy{group \textit{fluffy}}
\def\GroupPorous{group \textit{porous}}
\def\SolidGroup{\textit{solid} group}
\def\FluffyGroup{\textit{fluffy} group}
\def\PorousGroup{\textit{porous} group}

\def\SolidGrain{\mbox{SOLID\_1}}
\def\SolidAggregate{\mbox{SOLID\_2}}
\def\FluffyFractal{\mbox{FLUFFY\_1}}
\def\PorousAgglomerate{\mbox{POROUS\_1}}
\def\PorousAgglomerateCluster{\mbox{POROUS\_2}}
\def\PorousSolidRimParticle{\mbox{POROUS\_SOLID\_1}}
\def\PorousSolidFebo{\mbox{POROUS\_SOLID\_2}}
\def\FluffySolidMix{\mbox{FLUFFY\_SOLID\_1}}

\def\mum{$\mu$m}
\def\metersecond{$\mathrm{m\,s^ {-1}}$}
\def\kms{$\mathrm{km\,s^ {-1}}$}
\def\density{$\mathrm{kg\,m^ {-3}}$}

\title{Synthesis of the Morphological Description of\\Cometary Dust at Comet 67P}

\author{C. G{\"u}ttler \inst{\ref{inst:MPS}}
        \and T. Mannel \inst{\ref{inst:IWF},\ref{inst:GrazUni}}
        \and A. Rotundi \inst{\ref{inst:INAF_Rome},\ref{inst:Parthenope}}
        \and S. Merouane \inst{\ref{inst:MPS}}
        \and M. Fulle \inst{\ref{inst:INAF_Trieste}}
        \and D. Bockel{\'e}e-Morvan \inst{\ref{inst:LESIA}}
        \and J. Lasue \inst{\ref{inst:Toulouse}}
        \and A. C. Levasseur-Regourd \inst{\ref{inst:latmos_aclr}}
        \and J. Blum \inst{\ref{inst:IGEP}}
        \and G. Naletto \inst{\ref{inst:Naletto},\ref{inst:CISAS_PD},\ref{inst:CNR_PD}}
        \and H. Sierks \inst{\ref{inst:MPS}}
        \and M. Hilchenbach \inst{\ref{inst:MPS}}
        \and C. Tubiana \inst{\ref{inst:MPS}}
        \and F. Capaccioni \inst{\ref{inst:INAF_Rome}}
        \and J. A. Paquette \inst{\ref{inst:MPS}}
        \and A. Flandes \inst{\ref{inst:Flandes}}
        \and F. Moreno \inst{\ref{inst:IAF_Granada}}
        \and J. Agarwal \inst{\ref{inst:MPS}}
        \and D. Bodewits \inst{\ref{inst:auburn}}
        \and I. Bertini \inst{\ref{inst:UDP}}
        \and G. P. Tozzi \inst{\ref{inst:Tozzi}}
        \and K. Hornung \inst{\ref{inst:Hornung}}
        \and Y. Langevin \inst{\ref{inst:Langevin}}
        \and H. Kr{\"u}ger \inst{\ref{inst:MPS}}
        \and A. Longobardo \inst{\ref{inst:INAF_Rome}}
        \and V. Della Corte \inst{\ref{inst:INAF_Rome}}
        \and I. T{\'o}th \inst{\ref{inst:toth}}
        \and G. Filacchione \inst{\ref{inst:INAF_Rome}}
        \and S. L. Ivanovski \inst{\ref{inst:INAF_Trieste}}
        \and S. Mottola \inst{\ref{inst:DLR_Berlin}}
        \and G. Rinaldi \inst{\ref{inst:INAF_Rome}}
        }

\institute{Max Planck Institute for Solar System Research, Justus-von-Liebig-Weg 3, 37077 G{\"o}ttingen, Germany\label{inst:MPS}
           \and Space Research Institute, Austrian Academy of Sciences, Schmiedlstrasse 6, 8042 Graz, Austria\label{inst:IWF}
           \and Physics Institute, University of Graz, Universit{\"a}tsplatz 5, 8010 Graz, Austria\label{inst:GrazUni}
           \and INAF - Istituto di Astrofisica e Planetologia Spaziali, Via Fosso del Cavaliere 100, I-00133 Rome, Italy\label{inst:INAF_Rome}
           \and Universit{\'a} degli Studi di Napoli Parthenope, Dip. di Scienze e Tecnologie, CDN IC4, I-80143 Naples, Italy\label{inst:Parthenope}
           \and INAF - Osservatorio Astronomico, Via Tiepolo 11, I-34143 Trieste, Italy\label{inst:INAF_Trieste}
           \and LESIA, Observatoire de Paris, Universit{\'e} PSL, CNRS, Univ. Paris Diderot, Sorbonne Paris Cit\'e, Sorbonne Universit\'e, 5 Place J. Janssen, 92195 Meudon Pricipal Cedex, France\label{inst:LESIA}
           \and IRAP, Universit{\'e} de Toulouse, CNRS, UPS, CNES, Toulouse, France\label{inst:Toulouse}
           \and LATMOS, Sorbonne Univ., CNRS, UVSQ, Campus Pierre et Marie Curie, BC 102, 4 place Jussieu, 75005 Paris, France\label{inst:latmos_aclr}
           \and Institut f\"ur Geophysik und extraterrestrische Physik, Technische Universit\"at Braunschweig, Mendelssohnstr. 3, 38106 Braunschweig, Germany\label{inst:IGEP}
           \and University of Padova, Department of Physics and Astronomy ``Galileo Galilei'', Via Marzolo 8, 35131 Padova, Italy\label{inst:Naletto}
           \and University of Padova, Center of Studies and Activities for Space (CISAS) ``G. Colombo'', Via Venezia 15, 35131 Padova, Italy\label{inst:CISAS_PD}
           \and CNR-IFN UOS Padova LUXOR, Via Trasea 7, 35131 Padova, Italy\label{inst:CNR_PD}
           \and Ciencias Espaciales, Instituto de Geof{\'i}sica, Universidad Nacional Aut{\'o}noma de M{\'e}xico, Coyoac{\'a}n, Mexico City 04510, Mexico\label{inst:Flandes}
           \and Instituto  de Astrof{\'i}sica de Andaluc{\'i}a (CSIC), c/ Glorieta de la Astronomia s/n, 18008 Granada, Spain\label{inst:IAF_Granada}
           \and University of Padova, Department of Physics and Astronomy ``Galileo Galilei'', Vicolo dell'Osservatorio 3, 35122 Padova, Italy\label{inst:UDP}
           \and Osservatorio Astrofisico di Arcetri, INAF, Firenze, Italy\label{inst:Tozzi}
           \and Universit{\"a}t der Bundeswehr M{\"u}nchen, LRT-7, 85577 Neubiberg, Germany\label{inst:Hornung}
           \and Institut d'Astrophysique Spatiale, CNRS/Univ. Paris-Sud, F-91405 Orsay, France\label{inst:Langevin}
           \and Auburn University, Physics Department, 206 Allison Laboratory, Auburn, AL 36849, USA\label{inst:auburn}
           \and Konkoly Observatory, PO Box 67, 1525 Budapest, Hungary\label{inst:toth}
           \and Deutsches Zentrum f{\"u}r Luft- und Raumfahrt (DLR), Institut f{\"u}r Planetenforschung, Rutherfordstra{\ss}e 23, 12489 Berlin, Germany\label{inst:DLR_Berlin}
           }

\abstract{
	Before Rosetta, the space missions Giotto and Stardust shaped our view on cometary dust, supported by plentiful data from Earth based observations and interplanetary dust particles collected in the Earth's atmosphere.
	The Rosetta mission at comet 67P/Churyumov-Gerasimenko was equipped with a multitude of instruments designed to study cometary dust.
	While an abundant amount of data was presented in several individual papers, many focused on a dedicated measurement or topic.
  Different instruments, methods, and data sources provide different measurement parameters and potentially introduce different biases.
	This can be an advantage if the complementary aspect of such a complex data set can be exploited.
	However, it also poses a challenge in the comparison of results in the first place.
	The aim of this work therefore is to summarise dust results from Rosetta and before.
	We establish a simple classification as a common framework for inter-comparison.
	This classification is based on a dust particle's structure, porosity, and strength as well as its size.
	Depending on the instrumentation, these are not direct measurement parameters but we chose them as they were the most reliable to derive our model.
	The proposed classification already proved helpful in the Rosetta dust community and we propose to take it into consideration also beyond.
	In this manner we hope to better identify synergies between different instruments and methods in the future.
}

\keywords{comets: general -- comets: individual: 67P/Churyumov-Gerasimenko -- space vehicles: instruments} 
\begin{document}

\maketitle

\section{Introduction}
\label{sect:introduction}

When comets become active, they release gas and dust, where the latter is then carried away by the gas to form the cometary coma.
The detailed physical processes of the dust release from the surface are not well known.
However, given that cometary material is known to exhibit a very low strength \citep{AttreeEtal:2018, GroussinEtal:2015} and processes take place under the extremely low cometary gravity \citep{SierksEtal:2015}, the required forces for the dust lift-off are likely gentle.
This is the mechanism by which a comet -- formed 4.57 billion years ago -- slowly decomposes back into its building blocks.
The level of the primitiveness of these dust particles with respect to their formation time can be debated but it is clear that they still carry clues to the early formation of comets and our solar system.
It must be the ultimate goal of cometary dust studies -- whether from Earth or by space missions -- to interpret results from this viewpoint and aim to decipher these clues.

It was the purpose of Rosetta, ``ESA's Mission to the Origin of the Solar System'' \citep{SchulzEtal:2009}, to provide the data in support of this goal.
Three instruments on Rosetta were exclusively dedicated to the study of dust in the coma of comet 67P/Churyumov-Gerasimenko.
But several other instruments from the suite of 11 instruments on-board Rosetta and 10 instruments on the lander Philae were equally suited and successful in the study of cometary dust.
Results of these dust studies are presented in Sects. \ref{sect:MIDAS} to \ref{sect:Philae}.

Rosetta-era studies of cometary dust are standing on the shoulders of space missions like Giotto and Vega at 1P/Halley and Stardust at 81P/Wild 2.
Giotto was equipped with two dedicated dust instruments, the Dust Impact Detector System (DID) and the Particle Impact Analyzer (PIA), providing the first in-situ data of cometary dust shortly after its release from a comet.
Additionally, the Optical Probe Experiment (OPE) retrieved local dust brightness and polarisation.
The Vega spacecraft were equipped with the dust mass spectrometer PUMA \citep[on Vega 1;][]{KisselKrueger:1987, KruegerEtal:1991} and the dust particle detectors SP-1 and SP-2 \citep[on Vega 1 and 2, respectively;][]{ReinhardBattrick:1986}, all in-situ dust instruments.
The Stardust spacecraft, during its flyby at comet Wild 2, collected dust particles to bring them to Earth for detailed analysis (Stardust results are described further in Sect. \ref{sect:Stardust}).
Additional in-situ information was provided by Stardust's Dust Flux Monitor Instrument \citep[DFMI;][]{TuzzolinoEtAl:2004} and Cometary and Interstellar Dust Analyzer \citep[CIDA;][]{KisselEtal:2004}.

Aspects of cometary dust can also be studied from Earth:
Telescope observations can for instance determine levels of activity and the morphology of the large scale dust tails and trails, thus dynamics of dust particles;
photo-polarimetric studies allow the interpretation of the dust particles' structures.
Cometary dust particles, once lifted from a comet, can travel through the solar system and eventually cross the Earth orbit.
These are collected in the Earth stratosphere as interplanetary dust particles (IDPs) or on the Earth surface as micrometeorites (MMs).
These aspects will all be summarised and discussed in Sects. \ref{sect:IDP} and \ref{sect:EarthObs}.

The main goal of this work is to summarise Rosetta results on cometary dust and make these comparable among themselves but also with studies outside Rosetta.
Here we focus on the morphology and structure of cometary dust particles;
for the composition and mineralogy, the reader is referred to \citet[Rosetta]{EngrandEtal:2016}, \citet[Stardust]{ZolenskyEtal:2006} and others.
Many results were published by the different Rosetta instrument teams and due to the nature of the complementary instruments, measurement parameters are different and not directly comparable.
We will in Sect. \ref{sect:ClassificationMain} establish a clear language and classification for dust particles of different morphologies.
This is not a new definition but we try to summarise the consensus of the community and then rigorously stick to it.
Based on this, we will in Sect. \ref{sect:InstrumentResults} summarise results from Rosetta, Stardust, and Earth-based observations.
As a key result, these are summarised in Table \ref{tab:table} and Fig. \ref{fig:OverviewPlot}.
These are compared and discussed in Sect. \ref{sect:discussion} and a short conclusion is presented in Sect. \ref{sect:conclusion}.

This work is the result of a series of workshops and discussions, involving the largest part of the Rosetta dust community.
It is clear that due to the extent of data still being interpreted, this can only be a first step in a concerted understanding of Rosetta data in particular and cometary dust in general.
However, the work is ongoing and we aim to continue combining our results in the spirit of this paper.

\section{Classification}
\label{sect:ClassificationMain}

\subsection{General Nomenclature}
\label{sect:nomenclature}

Different communities or even different scientists tend to use slightly different nomenclatures.
This work is a large collaborative effort and the aim is to form a broad agreement (or at least identify disagreements).
It is therefore critical to be as explicit and precise as possible, which is why we provide here the used and agreed nomenclature.
We intend to keep consistence with the nomenclature used in the Stardust \citep[e.g.,][]{brownlee_comet_2006} as well as planetesimal formation \citep[e.g.,][]{DominikEtal:2007} communities.

A \textbf{grain} is the smallest component we consider in this study.
It is a solid particle with a tensile strength (typically > MPa) that is larger than forces acting in its environment.
A grain is likely irregular in shape but homogeneous in composition.
It is the constituent that forms the aggregates and agglomerates defined below.
Grains were created by condensation, either in the Solar System's protoplanetary disk or earlier in the interstellar medium or AGB-star outflows \citep[e.g.,][]{AlexanderEtal:2007}.
We do not specify the grain's material in this definition.

The term \textbf{monomer} is often used in this context and must not be mistaken with the definition of a monomer molecule.
In the dust community, monomer is used synonymously with grain, often in theoretic works.
We thus propose to keep this term but restrain its use to spherical or elliptical grains or their mathematical description.

We use the term (dense) \textbf{aggregate} for an intimate assemblage of grains, rigidly joined together and not readily dispersed.
Such dense aggregates might look like grains from the outside but in fact contain different mineralogical components (grains) in the inside.
The smallest components observed in the Stardust sample were these aggregates \citep{brownlee_comet_2006}.

A (porous) \textbf{agglomerate} is constituted of grains or dense aggregates.
The binding forces are much smaller than the grains' or aggregates' inner binding forces such that agglomerates are easily dispersed.
Agglomerates are the particles that are expected to form through dust agglomeration in the early protoplanetary disk \citep{DominikEtal:2007}.

The terms aggregate and agglomerate are often used synonymously, describing what we define here as agglomerate.
However, we see the need to formally discriminate between these two.
The precise distinction is not always consistent in the literature \citep[e.g.,][]{NicholsEtal:2002, Walter:2013} and we choose the definition that is prevalently used in the community whenever the two are discriminated at all.
We propose to address them as \emph{dense} aggregate and \emph{porous} agglomerates wherever the precise wording is important (as dense we consider porosities $< 10 \%$, see below).

Furthermore, we distinguish the case of a \textbf{fractal agglomerate}, which is showing a fractal and dendritic nature, implying a very high porosity of typically $> 99 \%$.
For these, the fractal dimension $D_\mathrm{f}$ defines the relation between mass $m$ and size $r$ as $m \propto r^ {D_\mathrm{f}}$ \citep[e.g., review by][]{Blum:2006}.
In our case, the relevant fractal agglomerates have $D_\mathrm{f} < 3$ and are typically in the range 1.5 -- 2.5.
This is consistent with particles formed by cluster-cluster agglomeration \citep{Blum:2006}.

Finally, we use the term \textbf{particle} as a generic term for any unspecified dust particle.
This can be anything from a monomer to an agglomerate and implies that the nature of a particle is not further known or considered.

\subsection{Structure and Porosity Classification}
\label{sect:StructurePorosityClassification}

\begin{figure*}[t]
  \centering
  \includegraphics[width=\textwidth]{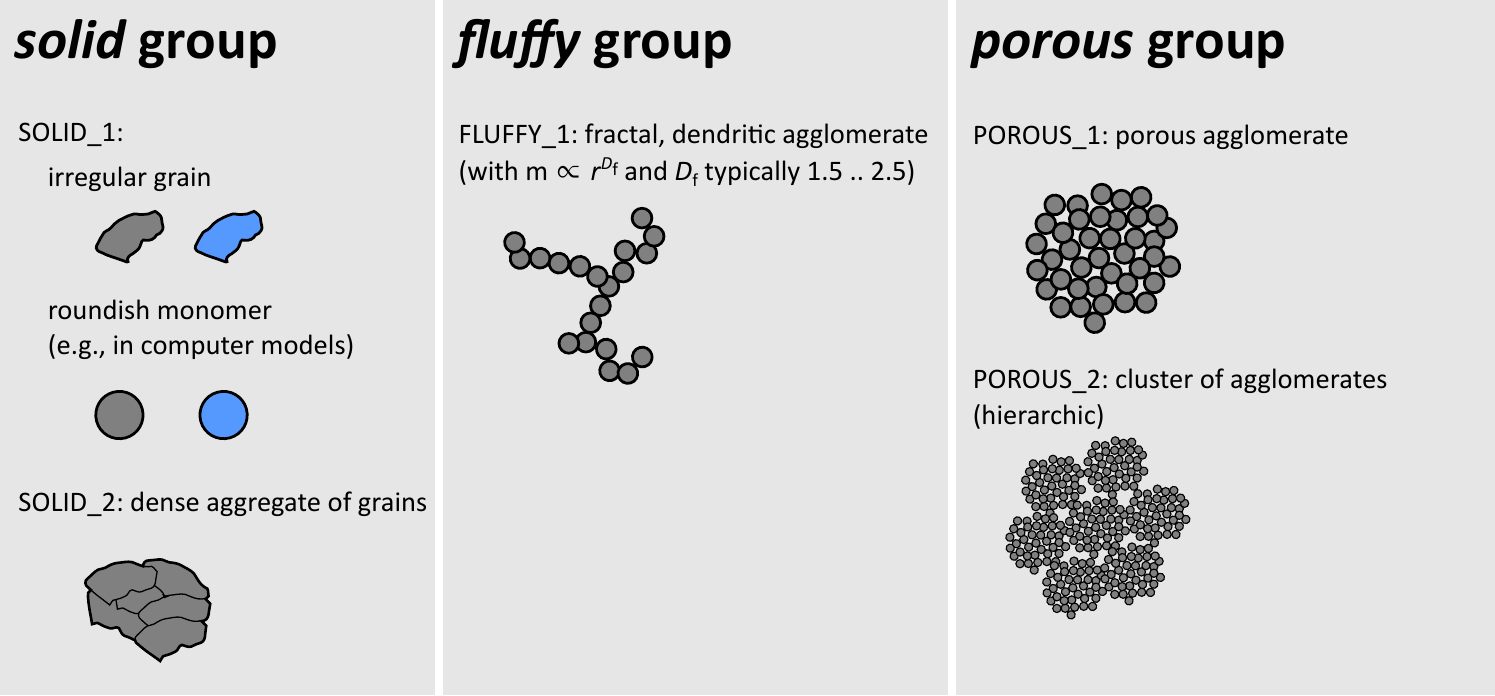}
  \caption{\label{fig:pictograms}Possible dust-particle structures, applying the three main groups defined in Sect. \ref{sect:StructurePorosityClassification}.
           The units of larger structures are drawn from circles for illustrative purposes only.
           In reality, these can be any of the \SolidGroup, where also colours (grey and blue) indicate that compositions can vary (e.g., ices).}
\end{figure*}

\begin{figure}[t]
  \centering
  \includegraphics[width=\columnwidth]{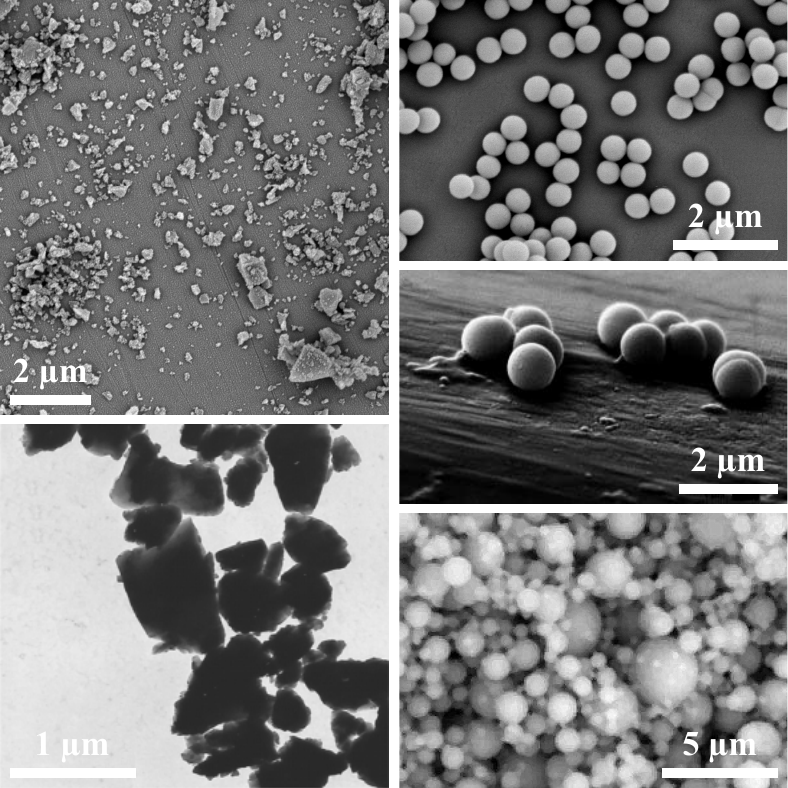}
  \caption{\label{fig:EXAMPLE_SOLID_1}Examples of laboratory analogues for the \SolidGrain\ types in Fig. \ref{fig:pictograms} (references in the text).}
\end{figure}

\begin{figure}[t]
  \centering
  \includegraphics[width=\columnwidth]{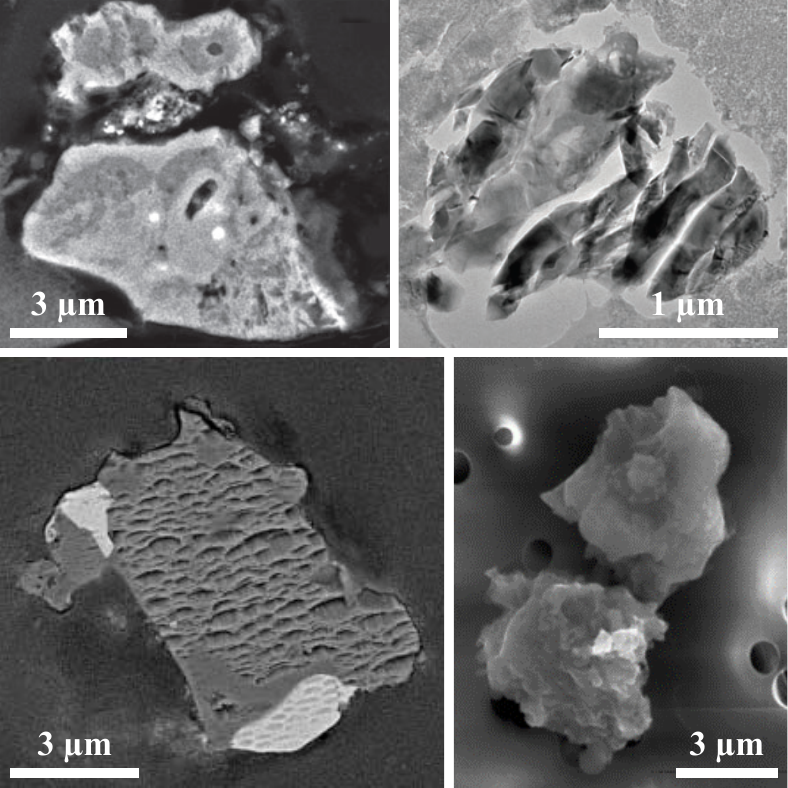}
  \caption{\label{fig:EXAMPLE_SOLID_2}Samples of the \SolidAggregate\ types in Fig. \ref{fig:pictograms} (references in the text).}
\end{figure}

\begin{figure}[t]
  \centering
  \includegraphics[width=\columnwidth]{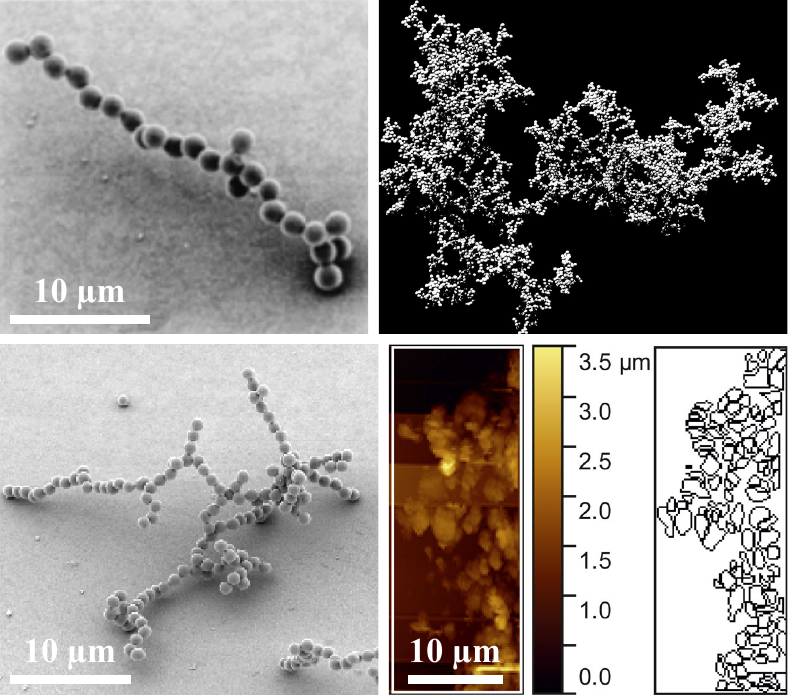}
  \caption{\label{fig:EXAMPLE_FRACTAL_1}Fractal particles from laboratory experiments (left), computer simulation (top right) and Rosetta/MIDAS (bottom right, left and right of scale bar; references in the text).
           The colour code and scale bar for the bottom right image denotes height.}
\end{figure}

\begin{figure}[t]
  \centering
  \includegraphics[width=\columnwidth]{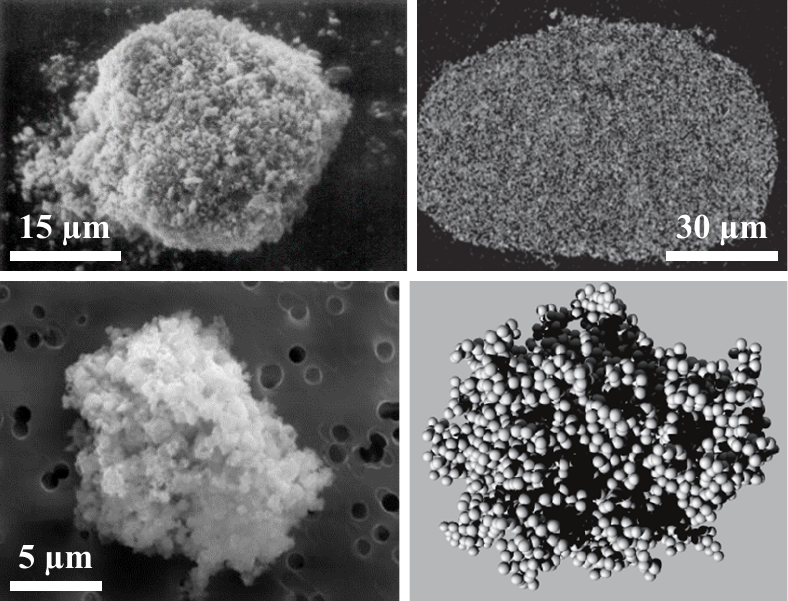}
  \caption{\label{fig:EXAMPLE_POROUS_1}Agglomerates from the \PorousAgglomerate\ type in Fig. \ref{fig:pictograms}.
  SEM image of a laboratory analogue agglomerate (top left), an IDP (bottom left), a tomographic cross section (top right), and a computer model (bottom right; references in the text).}
\end{figure}

\begin{figure}[t]
  \centering
  \includegraphics[width=\columnwidth]{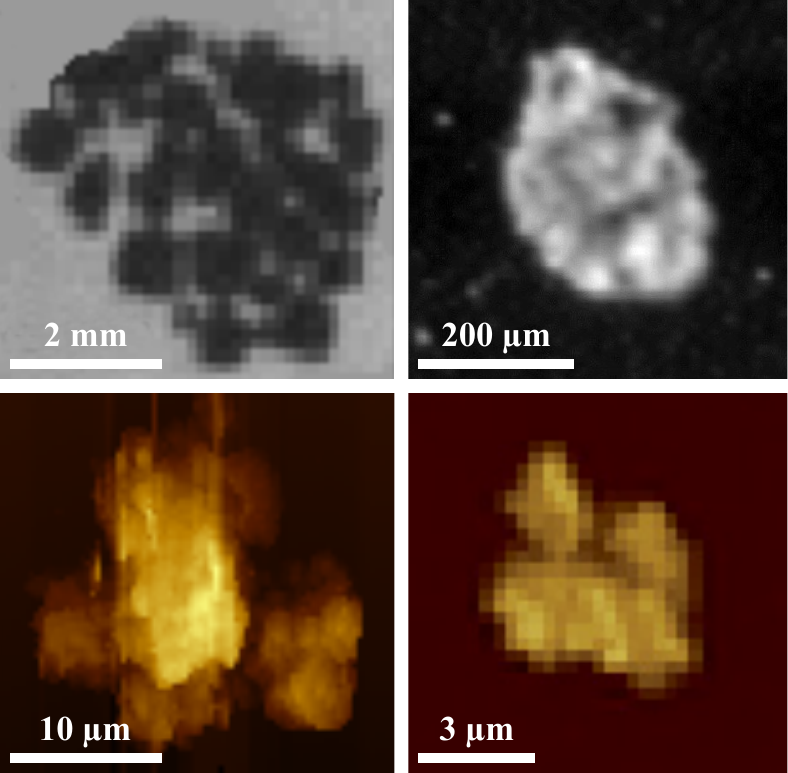}
  \caption{\label{fig:EXAMPLE_POROUS_2}Clusters of agglomerates from the \PorousAgglomerateCluster\ type in Fig. \ref{fig:pictograms}.
  Sample grown under microgravity (top left) and Rosetta/COSIMA (top right) and Rosetta/MIDAS samples (bottom; references in the text).}
\end{figure}

Based on the general nomenclature above, we further refine the description of the structures and porosities of dust particles.
Besides a particle's size, the porosity and structure are parameters which are to some degree accessible for Rosetta's dust instruments and are a focus of this work.
We introduce three groups, which will prove to be useful in terms of categorizing Rosetta dust observations summarised below.
Each group comprises physical properties and a structure, which can explain these properties.
Specifically, the three discriminating properties chosen here are (a) porosity, (b) structure, and (c) strength.
Various structures can be possible within these groups, which are illustrated in Fig. \ref{fig:pictograms} and explained below one by one.
Also, each of these structures will be compared to examples in nature, laboratory, or theory in Figs. \ref{fig:EXAMPLE_SOLID_1} -- \ref{fig:EXAMPLE_POROUS_2}.


The \textbf{\GroupSolid} describes particles with (a) a porosity < 10 \%, that are (b) consolidated and (c) exhibit a high strength similar to rock.
Particles that fall into this group are the grains and dense aggregates described in Sect. \ref{sect:nomenclature}, as well as chondrules or calcium aluminium rich inclusions (CAIs).
The tensile strength should be in the MPa range and higher, which is only the case for solid particles of low porosity.
The latter is chosen to be < 10 \%, to be much smaller than the random close packing of a granular medium \citep[$\sim 40$ \%,][]{OnodaLiniger:1990}, clearly discriminating between compressed agglomerates.
The most reasonable mechanism to create these low porosities for cometary particles is thermal processing, i.e., compaction through (partial) melting or vapour transport.

We identify two structures that fall into this group.
Irregular grains and spherical monomers (\SolidGrain\ in Fig. \ref{fig:pictograms}) are the smallest.
Examples of irregular grains used in laboratory analogue experiments are shown in Fig. \ref{fig:EXAMPLE_SOLID_1} (left).
Many different materials have been used in laboratory experiments, while the examples here are from diamond \citep[top left,][]{PoppeEtal:2000a} and Forsterite \citep[bottom left,][]{TamanaiEtal:2006}.
Spherical monomers can easily be formed in the laboratory from super saturated gas or liquid phases and are also used for analogue experiments in astrophysics.
All three examples in Fig. \ref{fig:EXAMPLE_SOLID_1} (right) are from SiO$_2$ but with different size distributions \citep[top to bottom:][]{PoppeEtal:2000a, ColangeliEtal:2003, BrissetEtal:2017}.
Monomers in nature are not perfectly spherical and exhibit surface roughness.

If grains form a dense aggregate, we expect a morphology like \SolidAggregate, which is an idealized (simplified) Stardust particle.
Figure \ref{fig:EXAMPLE_SOLID_2} shows three thin sections of solid aggregates from Stardust (top: \citet{Brownlee:2014}; bottom left: \citet{brownlee_comet_2006}).
Also some interplanetary dust particles (IDPs) collected in the Earth's stratosphere in the NASA Cosmic Dust Catalog\footnote{Example images of interplanetary dust particles in this work are from the NASA Cosmic Dust Catalog Volume 15 from 1997 \citep[see e.g.][]{Brownlee:1985, Brownlee:2016}.} resemble this morphology, an example is presented in Fig. \ref{fig:EXAMPLE_SOLID_2} (L2021B6, bottom right).

The structures following hereafter are based on agglomerated particles from the solid group.
\SolidGrain\ are drawn grey (e.g., silicates) or blue (e.g., ices) to make clear that the composition can be varying.
However, we want to leave the shape and composition of the constituent grains open on that scale and therefore assume that the agglomerates below can form out of any of those grains in any mixed state.
The composition of many agglomerates in Stardust and Rosetta is known but fall outside the scope of this paper.


The second class, \textbf{\GroupFluffy}, describes agglomerates, which (a) have a very high porosity (> 95 \%).
These (b) are likely fractal and dendritic agglomerates, the only reasonable explanation for extreme porosities, and (c) show a very low strength (Pa range).
A visualized example is \FluffyFractal\ in Fig. \ref{fig:pictograms}.
Fractal agglomerates are very well known from the literature, in particular in the context of early planet formation \citep{Blum:2006}.
The examples in Fig. \ref{fig:EXAMPLE_FRACTAL_1} (left) show fractal agglomerates from SiO$_2$, grown under laboratory conditions.
The top one is a small agglomerate out of 1.9 \mum\ monomers from \citet{HeimEtal:1999} while the lower one is significantly larger and has a fractal dimension of $D_\mathrm{f} \approx 1.8$ \citep[not measured for this specific agglomerate but for similar ones formed under the same condition,][]{Blum:2004}.
The example on the top right is a fractal agglomerate formed in a computer simulation by ballistic cluster-cluster agglomeration, consists of 8192 monomers and exhibits a fractal dimension of 1.99 \citep{WadaEtal:2008}.
The bottom right example is from the Rosetta/MIDAS experiment and will be discussed in detail in Sect. \ref{sect:MIDAS}.

Finally, the \textbf{\GroupPorous} collects the remaining parameter range with particles of (a) porosities between 10 and 95 \%.
These are (b) considered as loosely bound agglomerates with (c) an intermediate but rather low strength, typically in the order of 1 Pa to 100 kPa.
Laboratory analogue experiments demonstrated that in the case of silicates, this depends only mildly on composition \citep{BlumEtal:2006}.
Due to their higher stickiness in collisions \citep{GundlachBlum:2015}, ice agglomerates may form easier in the solar nebula.
However, their intrinsic cohesion (tensile strength) is very similar to that of silicates \citep{GundlachEtal:2018a} as long as the temperatures are low.
For temperatures above $\sim 140$ K, \mum-sized ice particles start to sinter on timescales shorter than $\sim 10^5$ s so that for cometary nuclei close to the ice-evaporation front, the mechanical strength might be increased \citep{GundlachEtal:2018b}.
Sintering can also occur for silicates \citep{Poppe:2003} and organics \citep{KouchiEtal:2002}.

Figure \ref{fig:pictograms} provides two examples for this group:
\PorousAgglomerate\ is a van-der-Waals agglomerate with a rather homogeneous structure, bound by surface forces.
Similar agglomerates are studied in laboratories and computer simulations.
In Fig. \ref{fig:EXAMPLE_POROUS_1}, we present an SEM image of a loose agglomerate consisting of 0.5 \mum\ solid Zirconium silicate particle \citep[top left,][]{BlumMuench:1993}, an IDP from the NASA Cosmic Dust Catalog (L2021A1, bottom left), an x-ray tomography reconstructed cut though an agglomerate from SiO$_2$ monomers \citep[top right,][]{KotheEtal:2013}, and an agglomerate used for numeric simulations \citep[bottom right,][]{WadaEtal:2011}.

The sub-structure might not be as homogeneous and the second example for the \PorousGroup\ in Fig. \ref{fig:pictograms} (\PorousAgglomerateCluster) represents a cluster consisting of smaller agglomerates with voids in-between.
Such hierarchic agglomerates were produced in laboratory experiments as shown in Fig. \ref{fig:EXAMPLE_POROUS_2} (top left).
This is a back-light illuminated agglomerate, grown from smaller (100 \mum) agglomerates under microgravity conditions \citep{BrissetEtal:2016}.
The other two examples are from Rosetta/COSIMA (centre) and Rosetta/MIDAS (right) and will be explained in detail in Sect. \ref{sect:MIDAS} and \ref{sect:COSIMA}.
A hierarchic agglomerate structure can formally be described as being fractal if the agglomerate consists of hierarchically structured (self-similar) sub-agglomerates.
In the case of \PorousAgglomerateCluster\ we assume either one single sub-agglomerate size or few cascades of sub-structures, such that all requirements for the \PorousGroup\ are still fulfilled.
The strength of these clusters is significantly smaller than \PorousAgglomerate\ structures if the number of contacts is small at sub-structure boundaries.

\begin{figure*}[t]
  \centering
  \includegraphics[width=\textwidth]{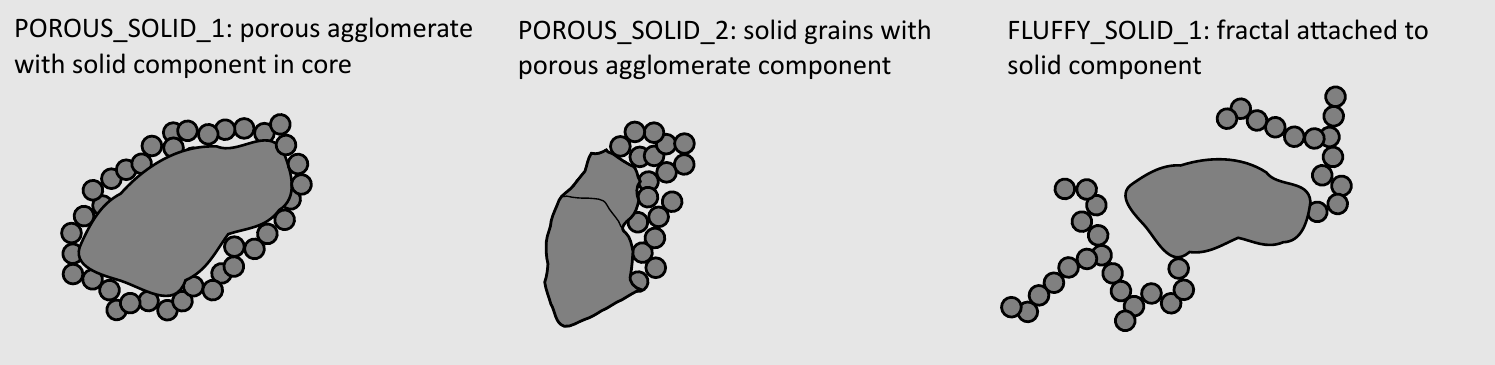}
  \caption{\label{fig:pictograms2}The classification from Sect. \ref{sect:StructurePorosityClassification} and Fig. \ref{fig:pictograms} is not always unambiguous.
           These examples of \textbf{mixed cases} show aggregate structures that would be classified one way or another depending on the method applied (e.g., surface microscopy vs. mass determination vs. light scattering)}
\end{figure*}

It should be noted that a classification by these three groups is not always unambiguous.
Structure, porosity, and strength have a likely but non-mandatory correlation.
It can therefore be possible that a studied particle shares properties of more than one group such that a classification is not easily possible.
This is in particular the case when also the structure shows properties of different groups as in the \textbf{mixed cases} in Fig. \ref{fig:pictograms2}.

\begin{figure}[t]
  \centering
  \includegraphics[width=\columnwidth]{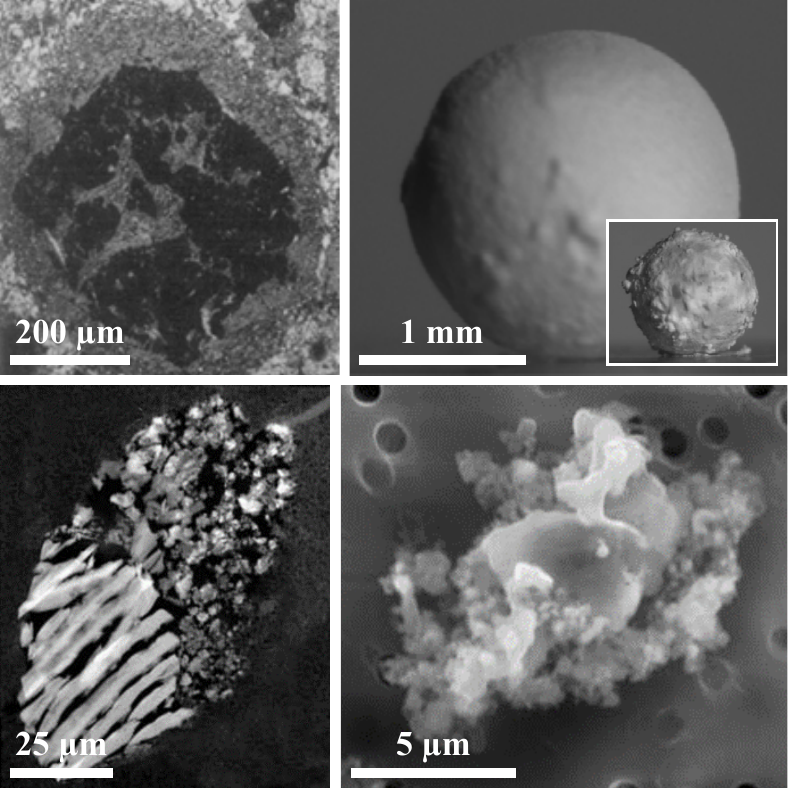}
  \caption{\label{fig:EXAMPLE_MIXES}Examples for mixed cases in Fig. \ref{fig:pictograms2}. (references in the text).}
\end{figure}

\PorousSolidRimParticle\ is a particle from the solid group, mantled with an agglomerate layer.
An example for such an agglomerate is the common picture of a mantled chondrule.
A polished cross-section of a rimmed chondrule by \citet{MetzlerEtal:1992} is shown on the top left of Fig. \ref{fig:EXAMPLE_MIXES} and an isolated chondrule analogue by \citet{BeitzEtal:2012} is shown in the top right (inset with different coating technique).
From a density measurement, one would interpret this particle as a member of the \SolidGroup, while the outer appearance (e.g., light scattering) would cloak it as a member of the \PorousGroup.
The Stardust particle T57 Febo \citep[also Fig. \ref{fig:EXAMPLE_MIXES} bottom left]{brownlee_comet_2006} is another mixed case \PorousSolidFebo.
Depending on the ratio between solid and mixed component, the group would be ambiguous.
Finally, a solid particle with an attached fractal structure as depicted in \FluffySolidMix\ was observed in IDPs (NASA Cosmic Dust Catalog, L2021A7, Fig. \ref{fig:EXAMPLE_MIXES} bottom right) and shares reflection and density properties from the \textit{solid} and \FluffyGroup.

Occurring ambiguities will further be discussed in Sec. \ref{sect:InstrumentResults} wherever they occur.

\section{State of Knowledge}
\label{sect:InstrumentResults}

In this section we summarise the knowledge on cometary dust with a focus on Rosetta.

Sections \ref{sect:MIDAS} through \ref{sect:Philae} focus on Rosetta dust instruments.
These are all different and thus complementary in nature:
The Micro-Imaging Dust Analysis System \citep[\textbf{MIDAS}; see][]{RiedlerEtal:2007} collected dust particles and determined their shape and structure with an atomic force microscope.
It was thus an \textit{in-situ} instrument with an \textit{imaging method}.
The same is the case for the Cometary Secondary Ion Mass Analyser \citep[\textbf{COSIMA}; see][]{KisselEtal:2007}, where collected particles were studied with a microscope and with a secondary ion mass spectrometer.
The Grain Impact Analyser and Dust Accumulator \citep[\textbf{GIADA}; see][]{ColangeliEtal:2007} was another \emph{in-situ} instrument but without an imaging method.
Particles were instead crossing a laser curtain and their size and speed was determined from the signal of \emph{scattered light} (Grain Detection System; GDS).
The particles then collided on the Impact Sensor (IS) where their momentum (thus mass if velocity is known) could be measured if they carried enough momentum.
The Optical, Spectroscopic, and Infrared Remote Imaging System \citep[\textbf{OSIRIS}; see][]{KellerEtal:2007}, consisting of a narrow- and a wide-angle camera, could observe -- within the inner coma -- individual (but still unresolved) dust particles as well as a diffuse signal from a large ensemble of undistinguishable particles.
In either case, the interpretation requires an assumption of the \emph{light scattering} properties.
The Visible and Infrared Thermal Imaging Spectrometer \citep[\textbf{VIRTIS}; see][]{CoradiniEtal:2007} could spectrally resolve the diffuse coma to infer colour and temperature of the unresolved dust.
As OSIRIS and GIADA above, also VIRTIS relies on the \emph{scattered light} and model assumptions on scattering properties.
On the surface of comet 67P, the lander Philae also studied the dust and we put the focus here on the dust monitor DIM as part of \textbf{Philae/SESAME} \citep{SeidenstickerEtal:2007} as well as the down-looking camera \textbf{Philae/ROLIS} \citep{MottolaEtal:2007} and the cameras for panoramic imaging \textbf{Philae/CIVA} \citep{BibringEtal:2007}.

Sections \ref{sect:Stardust} through \ref{sect:EarthObs} extend the picture beyond recent Rosetta findings.
We will consider constraints from the large body of Earth based observation as well as studies of cometary dust in laboratories, in particular the samples brought back to Earth by Stardust.

The intention of this section is to summarise the individual results and compile them into a comparative table (Table \ref{tab:table}).
While the summaries shall be descriptive and comprehensive, the resulting table is a simplification, which is intended as a framework and an aid to memory for cross comparison.
While individual instrument groups have so far interpreted specific instrument result, we are here aiming -- with all Rosetta instrument teams involved -- to homogenize our understanding and nomenclatures.
For a more general and complementary review of cometary dust with a focus on Rosetta, the reader is also referred to the article by \citet{LevasseurRegourdEtal:2018}.

\subsection{Rosetta/MIDAS}
\label{sect:MIDAS}

The MIDAS atomic force microscope revealed the surface structure of particles with nanometre resolution for 1 -- 50 \mum\ sized particles.
All studied particles show surfaces with textures that can be interpreted as that of an agglomerate consisting of smaller subunits, which could again be of agglomerate structure \citep{bentley_midas_2016}.


One particle was pointed out to exhibit an extraordinarily loose packing of subunits, and a sophisticated structural analysis was conducted \citep[Fig. \ref{fig:EXAMPLE_FRACTAL_1}, bottom right;][]{bentley_midas_2016, mannel_fractal_2016}.
Subunits range from 0.58 to 2.57 \mum\ with an average of 1.48 \mum\ \citep{bentley_midas_2016}, while it cannot be excluded that these subunits are again agglomerates with subunit sizes less than about 500 nm.
The particle is expected to have compacted during collection so that its image can be interpreted as a projection of the original structure onto the target plane (Fig. \ref{fig:EXAMPLE_FRACTAL_1}, bottom right) which was determined to be fractal with a fractal dimension $D_\mathrm{f} = 1.7 \pm 0.1$ \table\ \citep{mannel_fractal_2016}.
Only a $37 \times 20 \; \mathrm{\mu m}^2$ area of the particle was scanned and an attempt to scan the particle with adapted parameters showed that fragmentation destroyed the probably very fragile particle.
The representative size (disc of same area) of this particle is about 15 \mum\ diameter, while one could argue that the lateral extension is no larger than 80 \mum, which results in a representative size of 30 \mum\ diameter.
We therefore consider a range of 15 -- 30 \mum\ \table\ for the MIDAS \FluffyGroup.
It should be noted that since this was the only fractal collected by MIDAS, and the overall number of collected particles is unknown but assumed to be low, it is not feasible to determine a ratio of fractal versus non-fractal particles.

\begin{figure}[t]
  \centering
  \includegraphics[width=\columnwidth]{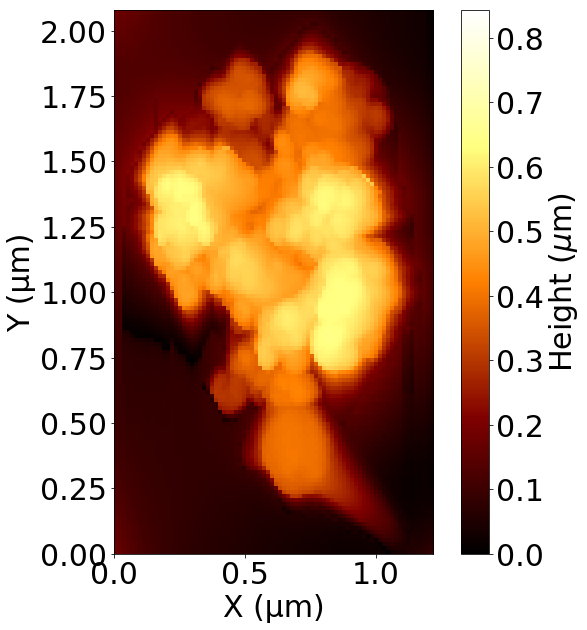}
  \caption{\label{fig:MidasSmallest}MIDAS image of an agglomerate particle sticking to the side of a tip that was acquired using a calibration target with sharp spikes \citep{MannelEtal:submitted}.
           The smooth round feature at the bottom is the tip apex and the straight line to the bottom right corner is a structure supporting the tip.
           The image has a pixel resolution of about 15 nm and was acquired on 2015-12-08.}
\end{figure}

All remaining particles with sizes about 10 -- 50 \mum\ show surface features in the order of 1 \mum, that are interpreted as loosely bound subunits (\GroupPorous\ in Sect. \ref{sect:StructurePorosityClassification}).
As an example, Fig. \ref{fig:EXAMPLE_POROUS_2} (bottom) shows two of these agglomerates in different sizes:
The left particle of $\sim20$ \mum\ diameter shows several sub-structures, of which the structure at the bottom right ($\lesssim10$ \mum) is the most prominent.
The particle in the bottom right of Fig. \ref{fig:EXAMPLE_POROUS_2} -- in the size of the previous subunit and measured with higher resolution -- again features about four subunits.
On the next smaller scale, Fig. \ref{fig:MidasSmallest} shows a 1 \mum\ particle that consists of a few hundred nanometre-sized subunits that again exhibit surface features possibly indicating that they are again agglomerates \citep{MannelEtal:submitted}.
Overall, this indicates that the 10 -- 50 \mum\ sized particles might have a hierarchical structure of agglomerates of agglomerates, resembling the example \PorousAgglomerateCluster.

The sizes of cavities compared to the subunits suggests a packing below maximal density, while, on the other hand, the density is certainly higher than that of the fractal particle described above.
Most of the 10 -- 50 \mum\ sized particles decomposed into many smaller fragments during scanning, which are of similar size as features observed on their surface.
This is suggesting that they are at least to a large amount consisting of subunits similar to those shaping their surface and are not mistaken as particles resembling the examples \PorousSolidRimParticle\ or \SolidAggregate.
The sizes of subunits are moreover similar to those of the fractal particle described by \citet{mannel_fractal_2016}.
This could be a sign for both, the fractal and the denser particles, having formed out of subunits from a similar reservoir and both being at least to a large extent agglomerates.

The smallest individual particles detected are between 1 and 10 \mum\ \citep[][and Fig. \ref{fig:EXAMPLE_POROUS_2} bottom right]{bentley_midas_2016}.
They are less numerous than larger particles and did not fragment during scanning, pointing towards a higher strength.
Their surfaces are similar to those of the 10 -- 50 \mum\ particles but in a size range of the large particles' subunits.
As the 1 -- 10 \mum\ particles were scanned with higher resolutions, it is possible to resolve features of sizes down to about 500 nm on their surfaces (see Fig. \ref{fig:MidasSmallest}, discussed below), and highest resolution scans (8 and 15 nm/pixel) even resolve features with sizes of about 100 nm.
The deep trenches observed between the few hundred nanometres features indicate that the 1 -- 10 \mum\ particles might be agglomerates rather than members of the \SolidAggregate\ class.
However, it should be mentioned that they were never seen to disintegrate and thus they could also have a solid core covered with subunits like \PorousSolidRimParticle.
Relying on the deep trenching we suggest to classify 1 -- 10 \mum\ particles as agglomerates with a non-negligible porosity, which overall puts MIDAS dust particles of 1 -- 50 \mum\ size into the \PorousGroup.


Particles matching the \SolidGroup\ were not strictly observed by MIDAS.
Neither a porosity, nor the inner structure of the dust can be determined with the current data.
Also the size of the smallest refractory subunits cannot be determined due to the resolution limit of MIDAS and the ambiguity between surface features created by roughness or subunits.
The smallest identified features have sizes between 50 -- 200 nm \citep[Fig. \ref{fig:MidasSmallest} and][]{MannelEtal:submitted}.
They were detected in a special imaging mode where the instrument picked up particles or particle fragments which could have altered the particles or fragments.
It is expected that particle alteration first alters the arrangement of the subunits, and a higher stress is required to change subunit properties.
As the subunit size distributions of rather porous looking picked up particles and more compressed looking particles are similar, we expect that no major alteration of subunit sizes occurred.
However, if these smallest 50 -- 200 nm sized features are only surface features of consolidated larger units, then the next larger units of less than about 500 nm size would be the candidates for the smallest solid unit detected by MIDAS.
To conclude, with a significant uncertainty, we consider particles of 50 -- 500 nm \table\ as the smallest particles, possibly classified into our \SolidGroup.

\begin{table*}[t]
  \caption[caption]{\label{tab:table}Summary of Rosetta and Stardust classification.
           The table collects mostly sizes (all in diameter) for inter-comparison and classifications into morphological groups following Sect. \ref{sect:StructurePorosityClassification}.
           A visual representation of this table is presented in Fig. \ref{fig:OverviewPlot}.\\
           \textbf{Note:} The terminology used in particular for Stardust is described in detail in Sect. \ref{sect:Stardust} (see also Fig. \ref{fig:stardust_figure}).}
  \centering \footnotesize
  \begin{tabular}{|l|c|c|c|c|c|c|}
    \hline
                          & \textbf{MIDAS}           & \textbf{COSIMA} & \textbf{GIADA}      & \textbf{OSIRIS}        & \textbf{VIRTIS}     & \textbf{Stardust}\\
    \hline
    \textbf{\PorousGroup} & 1 -- 50 \mum             & 14 -- 300 \mum  & 0.1 -- 0.8 mm       & $\sim$100 \mum\ -- 1 m & dominating size     & particle creating\\
    - porosity 10-95 \%   &                          & on target;      &                     & dominant               & distribution        & track A with\\
    - aggregate           &                          & up to           &                     & scatterers             & (diff. slope        & multiple terminals\\
    - low strength        &                          & mm-range        &                     &                        & --2.5 to --3)       & or track B\\
                          &                          & parents         &                     &                        &                     & 1 -- 100 \mum\\
    \hline
    \textbf{\FluffyGroup} & fractal: 15 -- 30 \mum\  & no indication   & 0.1 -- 10 mm        & not dominant           & not excluded,       & particle creating bulbous\\
    - porosity >95 \%     & $D_\mathrm{f}=1.7\pm0.1$ &                 & $D_\mathrm{f}<1.9$, & scatterers             & consistent with     & tracks (B for coupled,\\
    - likely fractal      & constituent              &                 & $\sim$23 \% of GDS  &                        & moderate super-     & A* or C for fluffy GIADA\\
    - very low strength   & particles:               &                 & detections          &                        & heating in normal   & detections), aluminium foil\\
                          & < 1.5 \mum               &                 &                     &                        & activity            & clusters. Up to 100 \mum \\
    \hline
    \textbf{\SolidGroup}  & 50 -- 500 nm             & CAI candidate   & 0.15 -- 0.5 mm      & no indication          & outburst:           & particle creating\\
    - porosity <10 \%     & fragments                & and specular    & $\sim$4000 \density &                        & temperature         & track A with single\\
    - consolidated        & collected on tip         & reflection      &                     &                        & requires            & or multiple terminals,\\
    - high strength       &                          & 5 -- 15 \mum    &                     &                        & 0.1 \mum\ particles & 10s of nm, 1 -- 100 \mum\\
    \hline
  \end{tabular}
\end{table*}

\subsection{Rosetta/COSIMA}
\label{sect:COSIMA}

The COSIMA instrument collected dust particles to image these with an optical microscope (COSISCOPE) and then perform secondary ion mass spectroscopy (SIMS).
Observed dust particles range from COSISCOPE's resolution limit of 14 \mum\ to around one millimetre \table.
During collection, the particles collided with the targets with a varying relative velocity around few \metersecond\ with respect to the Rosetta spacecraft \citep{RotundiEtal:2015}.
Upon impact they therefore fragmented and the adhering fragments show a power law size distribution with a differential power law exponent of $-3.1$ in average \citep{HilchenbachEtal:2016, MerouaneEtal:2016}.

The fragments were initially classified as \emph{compact particles}, \emph{rubble piles}, \emph{shattered clusters}, and \emph{glued cluster} \citep{LangevinEtal:2016, MerouaneEtal:2016}.
It was later shown that also \emph{compact particles} are fragile enough to be broken by mechanical pressure as well as by charge-up in the SIMS ion beam \citep{HilchenbachEtal:2017}.
\citet{EllerbroekEtal:2017} furthermore showed in laboratory analogue experiments that the four morphological groups defined by \citet{LangevinEtal:2016} can be explained through a variation of the collection velocity.
Based on this and the evidence that all particles can be further broken, we consider all as agglomerates according to the classification in Sect. \ref{sect:StructurePorosityClassification}.
Many particles show sub-structures, even down to the resolution limit of 14 \mum/pixel, which indicates that they are clusters consisting of smaller components, possibly again agglomerates (similar to \PorousAgglomerateCluster\ in Fig. \ref{fig:pictograms}).

In an attempt to infer mechanical properties from the impact fragmentation, \citet{HornungEtal:2016} used this picture for the impacting agglomerates and inferred a strength (they intentionally do not distinguish between tensile and shear strength) for the initially un-fragmented agglomerate.
In their model, the strength is determined by the binding force between sub-agglomerate structures, thus depends on their sizes, and they arrive at around 1000 Pa \table\ when assuming subunits of 10 -- 40 \mum, i.e., in the size around COSISCOPE's resolution limit.

Observations of sub-structures of non-fragmented agglomerates were interpreted by \citeauthor{HornungEtal:2016} as macroscopic filling factors defined as 1 -- porosity (on a 60 -- 300 \mum\ scale) in the range of 0.4 to 0.6.
These sub-structures were in turn assumed to be porous with the smallest solid unit of $\sim 0.2$ \mum\ diameter.
This is formally described by their size dependent filling factor $\phi \propto r^{-0.4}$, which implies a hierarchical cascade of sub-structures and porosities on all scales down to the solid grain.
Since the fragmentation model constrains strength boundaries rather than void spaces, the porosity of above 90 \% quoted by \citeauthor{HornungEtal:2016} is likely not a very strong restriction.
Based on a hierarchical porosity model for the dust agglomerates, the bulk of the dust particle porosity budget would reside in the size range below 14 \mum.
The exact values rest on the known and the adopted model parameters.

The agglomerates collected by COSIMA were mostly ice free as the instrument was kept warm inside and particles were studied days to weeks after collection.
However, they show a surprisingly large reflectance factor in the 3 -- 22 \% range \citep{LangevinEtal:2016, LangevinEtal:2017}.
Agglomerates were illuminated with a red LED (640 nm) at phase angles between 72 and 84 deg from two directions (roughly opposite, left and right in the image plane, cf. \citet[][Fig. 1]{LangevinEtal:2017}), one after another.
The measured reflectance, in particular the comparison between left and right illumination, was explained with scattering centres inside the agglomerate volume and an optical mean free path in the 20 -- 25 \mum\ range.
The required porosity depends on the size of the scattering centres and is estimated to be in the 50 -- 90 \% range \citep{LangevinEtal:2017}.

The COSIMA results can also provide a clue about the \SolidGroup:
\citet{PaquetteEtal:2016} have found compositions, which are consistent with CAI material.
These were discovered on separated spots in a 30 \mum\ rastered line scan of particle David Toisvesi.2.
A possible interpretation would be two or more solid CAI particles (thus \SolidGrain) of < 30 \mum, embedded in an agglomerate of porous nature.
Another hint for solid components is provided by specular reflections.
\citet{LangevinEtal:2017} interpret these as being produced by 5 -- 15 \mum\ crystalline facets.
This would be a grain of type \SolidGrain.
The size estimate is derived from a comparison to reference olivine particles, which were dispersed on one of the flight targets before launch \citep{LangevinEtal:2017}.
Since the size constraint from the CAI candidate above is a weaker one, we consider COSIMA solid particles (\SolidGrain\ or \SolidAggregate) in the 5 -- 15 \mum\ range \table.

\subsection{Rosetta/GIADA}
\label{sect:GIADA}

GIADA measured the scattered light, speed, and momentum of dust particles \citep{dellacorte_shining_2015}.
On the basis of these data, it is possible to infer a dust particle's density dependent on its true shape, composition and micro-porosity \citep{fulle_comet_2016}.

The measurement range of GIADA fell between 0.3 -- 100 \metersecond\ in speed (higher velocities up to 300 \metersecond\ are less reliable) and $10^{-10} - 4 \cdot 10^{-4} \; \mathrm{kg \, m \, s^{-1}}$ in momentum, which results in masses between $1.0\cdot 10^{-12} - 1.3 \cdot 10^{-3}$ kg.
The particle equivalent diameters fall between 60 -- 200 \mum\ for high albedo material (kaolinite) and 150 -- 800 \mum\ for low albedo material \citep[amorphous carbon;][]{DellaCorteEtal:2016}.

All particles measured with GIADA GDS and IS show densities enveloped by dust bulk densities of Fe-sulphides (4600 \density) and hydrocarbons (1200 \density) where either an oblate or prolate ellipsoidal shape with aspect ratios up to 10 was assumed \citep{FulleEtal:2017}.
The mean dust bulk density results in $785^{+520}_{-115}$ \density, where the large uncertainty arises from the unknown shape.
However, an average spherical shape is in good agreement with the prolate and oblate ellipsoids framing the whole dataset (except for a few outliers).
\citet{FulleEtal:2017} inferred the dust volume filling factor to be $0.59 \pm 0.08$.
With this, the majority of the dust detected by GIADA is described to be porous agglomerates (\GroupPorous).
Their sizes span the whole detection range of GIADA from roughly 0.15 to 0.8 mm \table.

Two dust populations drop out of this characterisation:
First, \citet{FulleEtal:2017} presented particles with a density of around or above 4000 \density\ (cluster with small cross sections in their Fig. 1; for those, the assumed albedo was that of carbon, otherwise the density would be even higher).
These densities can only be explained by a compaction mechanism, which has to affect their strengths, making a loose agglomerate structure unlikely.
Their sizes are typically < 0.5 mm \table\ and, according to the classification in Sect. \ref{sect:StructurePorosityClassification}, they fall into the \SolidGroup.
It cannot be excluded that the structure resembles for instance \PorousSolidRimParticle\ in Fig. \ref{fig:pictograms2} if the solid core is large enough.
In any case the measurements demand the existence of macroscopic particles from the \SolidGroup.

The other extreme are dust particles that are inferred to be very low density, fluffy agglomerates \citep{dellacorte_shining_2015}.
In particular their densities and speeds were so low that these particles had insufficient momentum to produce a signal at the impact sensor.
They were only detected by the optical detection measurement sub-system (GDS) as showers of many small dust particles caused by large, low-density parent agglomerates fragmenting directly in front of the instrument \citep{fulle_density_2015}, where the confinement of the shower restricted the fragmentation to happen in close proximity.
The low speed and the fragmentation in front of the spacecraft were explained by \citeauthor{fulle_density_2015} with electrostatic forces:
The fluffy agglomerates got charged in the coma, which led to their disruption if the strength of the electrostatic field got larger than the strength keeping the agglomerate together.
\citeauthor{fulle_density_2015} estimated charge, mass, and cross-section of the parent agglomerates and derived an upper limit for the equivalent bulk density of less than 1 \density.
Moreover, the velocity of these fluffy agglomerates was smaller than the escape velocity from the comet.
Assuming that this velocity difference was caused by electrostatic deceleration of the particles at the spacecraft, \citet{FulleBlum:2017} determined the size of the fluffy agglomerates before breakup to be in the millimetre range.
Since the only plausible way to grow such a large and porous particle is fractal growth, \citet{fulle_unexpected_2016} calculated a fractal dimension of $D_\mathrm{f} \approx 1.87$ \table\ for the fluffy agglomerates.
The GDS showers, i.e. the fluffy particles, were often accompanied by the signal of a compact particle at the IS (GDS+IS detection), which is interpreted as the fluffy particles being attached to a compact one until disruption shortly before detection. 
It is not known if all fluffy particles were attached to a compact particle as the cross sections of the GDS showers were larger than GIADA's entrance area, leading to the possibility of the compact particle not entering the instrument and escaping detection.
Furthermore, GDS detections of single particles with low speeds and not enough momentum to create an IS signal could also have been caused by fluffy particles, although a determination of density, and thus clear assignment to the \GroupFluffy\ or \GroupPorous\ is not possible. 

The chance of a particle being a fluffy agglomerate is about 23 \% \table\ \citet{FulleBlum:2017} if all GDS single detections are counted as \GroupPorous, or 58 \% if the latter are counted as \GroupFluffy.

\subsection{Rosetta/OSIRIS}

With the Rosetta/OSIRIS camera system, dust in the coma of 67P could be studied remotely, thus non invasively, through observations in different colour filters of the solar light they scatter.
We have to distinguish between individual particles, which were in most cases unresolved (smaller than one pixel), and a diffuse signal from a large ensemble of undistinguishable particles.

Individual particles were first described by \citet{RotundiEtal:2015}, who determined a dust-particle size distribution, later analysed in its time evolution by \citet{FulleEtal:2016} and \citet{OttEtal:2017}.
Detectable sizes by this method are typically in the range of centimetres and decimetres.
\citet{AgarwalEtal:2016} studied the larger end of particles observed with OSIRIS, which were close to the comet and far from the spacecraft, with the largest about 80 cm diameter\footnote{The sizes shown in the paper are valid for particles that have a phase function and albedo as described by \citet{KolokolovaEtal:2004}, contrary to what is stated in the paper text. If they show reflection properties like the nucleus \citep{GuettlerEtal:2017}, the sizes should be corrected up (increased) by a factor 4.4 (Agarwal pers. comm.).} \table.
Only one particle was detected that is resolved by the cameras \citep[i.e., larger than 1 pixel, Fig. 7 in][]{FulleEtal:2016}, where the size is largely uncertain but likely larger than a metre.
\citet{FrattinEtal:2017} studied individual particles in different OSIRIS colour filters to assess their composition and, depending on their spectral slope, associated different particles with organics, silicates, or water ice.
All of these particles are expected to fall into the \PorousGroup\ as defined in Sect. \ref{sect:StructurePorosityClassification}.
They are too large to be fractals or grains and it is unlikely for them to be solid.

The smallest individual particles in OSIRIS were observed by \citet{GuettlerEtal:2017}.
These were close to the spacecraft (1 -- 100 metres) and the smallest measured 0.3 mm in diameter \table.
The sizes could have been smaller, depending on the scattering properties of the particles (see discussion in \citeauthor{GuettlerEtal:2017} and \citet{FulleEtal:2018}).
Density assumptions in the 100 -- 1000 \density\ range can explain the particles' observed acceleration, either through a rocket force (\citeauthor{GuettlerEtal:2017}) or by pure solar radiation pressure (\citeauthor{FulleEtal:2018}).
Based on this density, the agglomerates fall into the \PorousGroup\ \table.

The diffuse coma observed under different phase angle conditions was studied by \citet{BertiniEtal:2017, BertiniEtal:2019}.
With the comet outside the OSIRIS field of view (FoV; preferably by 90 deg), the Rosetta spacecraft was rotated around a vector perpendicular to the Sun direction and inside the comet-spacecraft-Sun plane to take images of the coma at a wide range of phase angles.
From the overall flux in the images (after filtering cosmics and individual dust particles), they computed a phase curve, which interestingly shows a concave ''smile shape'', featuring an absolute minimum at around 90 deg phase angle (see their Figs. 2 -- 4).

\citet{MorenoEtal:2018} succeeded in modelling the full phase function using elongated particles of diameter $\gtrsim 20$ \mum, which need to be aligned along the solar radiation direction.
In a complementary modelling attempt, \citet{MarkkanenEtal:2018} could reproduce the OSIRIS phase function at different times using aggregates in the 5 -- 100 \mum\ size range, consisting of sub-micrometre-sized organic grains and micrometre-sized silicate grains.
Indication for macroscopic particles (in contrast to dispersed sub-micrometre monomers) is also provided from laboratory analogous experiments by \citet{MunozEtal:2017}.
Overall, there are indications for particles smaller that the best OSIRIS resolution in OSIRIS data but interpretation and detailed studies are still ongoing.


\subsection{Rosetta/VIRTIS}
\label{sect:VIRTIS}

The Rosetta/VIRTIS dual channel spectrometer \citep{CoradiniEtal:2007} consisted of two instruments:
A point spectrometer VIRTIS-H (operating in the 2 -- 5 \mum\ spectral range with an FoV of $0.033^\circ \times 0.10^\circ$) and a line scanning imaging spectrometer VIRTIS-M (operating in the 0.25 -- 5 \mum\ spectral range with an FoV of $3.6^\circ \times 3.6^\circ$).
Due to its low spatial resolution and relatively long integration times (compared to Rosetta/OSIRIS), it could not study individual dust particles.
The strength of VIRTIS was the high spectral resolution and the extended wavelength range covering  thermal radiation in the 3 -- 5 \mum\ spectral range.
Spectra of the diffuse coma can be modelled to provide remote-sensing information, complementary to other Rosetta instruments.

A comprehensive study of the diffuse coma and outbursts observed with the VIRTIS-H channel on 13 and 14 Sep 2015 was presented by \citet{Bockelee-MorvanEtal:2017, Bockelee-MorvanEtal:2017erratum} and \citet{RinaldiEtal:2018}.
The two key results that we are here picking up are the particle size distribution in the quiescent coma and the detection of high-temperature grains (see below) during outburst.

\citet{Bockelee-MorvanEtal:2017} modelled the 2 -- 5 \mum\ infra-red emission of a collection of porous and fractal particles with Mie and Rayleigh-Gan-Debye theories (see \citeauthor{Bockelee-MorvanEtal:2017} for details), in order to explain the 2-\mum\ colour, colour temperature, and bolometric albedo measured on the spectra.
The best fit for the quiescent coma was achieved with a differential power index $\beta$ of the $n(a) \propto a^ {-\beta}$ size distribution in the range 2.5 -- 3 \table, consistent with the power index determined by other instruments \citep{RotundiEtal:2015, FulleEtal:2016}.
The observed 20 \% excess in colour temperature with respect to the equilibrium temperature can be attributed either to the presence of sub-micrometre particles made of absorbing material or, alternatively, to fractal agglomerates with sub-micrometre units.
The ratio of fractal versus porous agglomerates influences the minimum size of the particles in the size distribution fitting the measurements.
For a relative number of fractal agglomerates of 25 \% \citep{FulleBlum:2017}, particles at sizes below $\sim 20 - 30$ \mum\ should be under-abundant \citep{Bockelee-MorvanEtal:2017erratum}.
The scattering and thermal properties of 67P's diffuse coma are in line with the mean of values measured for moderately active comets, showing that 67P is not atypical concerning dust properties (\citeauthor{Bockelee-MorvanEtal:2017}).
    
The material detected shortly after outburst onset showed blue colours and colour temperatures as high as 550 and 650 K (Figs. 4 and 5 in \citet{Bockelee-MorvanEtal:2017}, and \citet{RinaldiEtal:2018}).
This was attributed to super-heating of very small particles, which got warmed by solar irradiation but could not sufficiently cool through infra-red emission.
The required particle size to explain the two properties is $\sim$ 0.1 \mum, and particles are believed to be individual, i.e., not bound in larger aggregates (see discussion in \citet{Bockelee-MorvanEtal:2017} and \citet{RinaldiEtal:2018}).
Since nanometre-sized particles may not be present in the general background coma, \citeauthor{Bockelee-MorvanEtal:2017} suggest that the outburst was disintegrating loosely bound agglomerates, which were otherwise bound by strong cohesion.
These smallest particles could fall into our \SolidGroup, although their strength and porosity is not constrained by VIRTIS.

\subsection{Rosetta/Philae}
\label{sect:Philae}

\subsubsection{DIM}
The Dust Impact Monitor (DIM) on board Philae was a 7 cm side cube designed to detect sub-millimetre- and millimetre-sized dust particles emitted from the nucleus of the comet employing piezoelectric detectors.
The cube had three active sensor sides, and each side had a total sensitive area of 24 cm$^2$ \citep{SeidenstickerEtal:2007}.
During the descent of Philae to the surface of 67P, DIM recorded an impact of a cometary dust particle (among many other impact signals identified as false impacts) at 2.4 km from the comet surface \citep{HirnEtal:2016, KruegerEtal:2015, FlandesEtal:2018}.
Experiments support the identification of this particle (aerogel was used as a comet analogue material to characterise the properties of this particle).
They are consistent with a particle radius of 0.9 mm, density of 250 \density, and porosity close to 90 \%.
Data and estimations also indicate that the particle likely moved at near 4 \metersecond\ with respect to the comet \citep{PodolakEtal:2016, FlandesEtal:2018}.

\subsubsection{ROLIS}
The Rosetta Lander Imaging System (ROLIS) performed observation of the original Philae landing site Agilkia during descent and later of Philae's final rest location Abydos.
In the Agilkia region, surface regolith was observed with a best resolution of 0.95 cm/pixel \citep{MottolaEtal:2015}.
Besides the power law size distribution of particles, the images reveal that small, decimetre-sized boulders show surface textures down to the resolution limit.
At the Abydos site, the material is more lumpy and no individual particles or pores can be distinguished at the resolution limit of 0.8 cm/pixel \citep{SchroederEtal:2017}.
Dust particles crossing the camera field of view were observed few centimetres as well as several metres from the camera \citep[Fig. 3 and Fig. 7, respectively, in][]{SchroederEtal:2017}.
No size or morphological information can be determined though.

\subsubsection{CIVA}
The Comet Infrared and Visible Analyser (CIVA) performed successful observations at the Abydos final landing site.
With a best resolution of 0.6 mm/pixel, the observed surface is in parts interpreted as pebbles with a dominating size of 5 -- 12 mm \citep{PouletEtal:2016}.
It should be noted that these are not clearly detached from the surface, which means that the observations are consistent with ROLIS and the interpretation is different.
CIVA has observed one isolated signal in the coma, which was interpreted as a detached particle by \citet[supplement Fig. S5]{BibringEtal:2015}, consistent with ROLIS observations.
Also here, no further properties of this particle candidate can be determined.

\subsection{Stardust Sample Collection}
\label{sect:Stardust}

The Stardust mission collected and returned cometary dust samples which are the only cometary samples of known origin available on Earth \citep{brownlee_comet_2006}.
The spacecraft made a fly-by at the Jupiter family comet 81P/Wild 2 in January 2004 at 234 km closest distance and 6.1 \kms\ relative velocity.
It captured more than 10,000 dust particles between 1 to 100 \mum\ in collectors of 3 cm thick silica aerogel tiles.
In addition, the aluminium frame around the aerogel tiles shows impact craters with residues of the particles.

All particles suffered alteration due to the capture, dominantly through heating to temperatures above the melting point of silica.
Larger particles over a micrometre in size are often reasonably well preserved due to their higher thermal inertia, whilst sub-micrometre dust was only able to survive when shielded by a larger particle \citep{brownlee_comet_2006, Rietmeijer:2016}.

\subsubsection{Aerogel Tracks}
\label{sect:StardustAero}

\begin{figure}[t]
  \centering
  \includegraphics[width=\columnwidth]{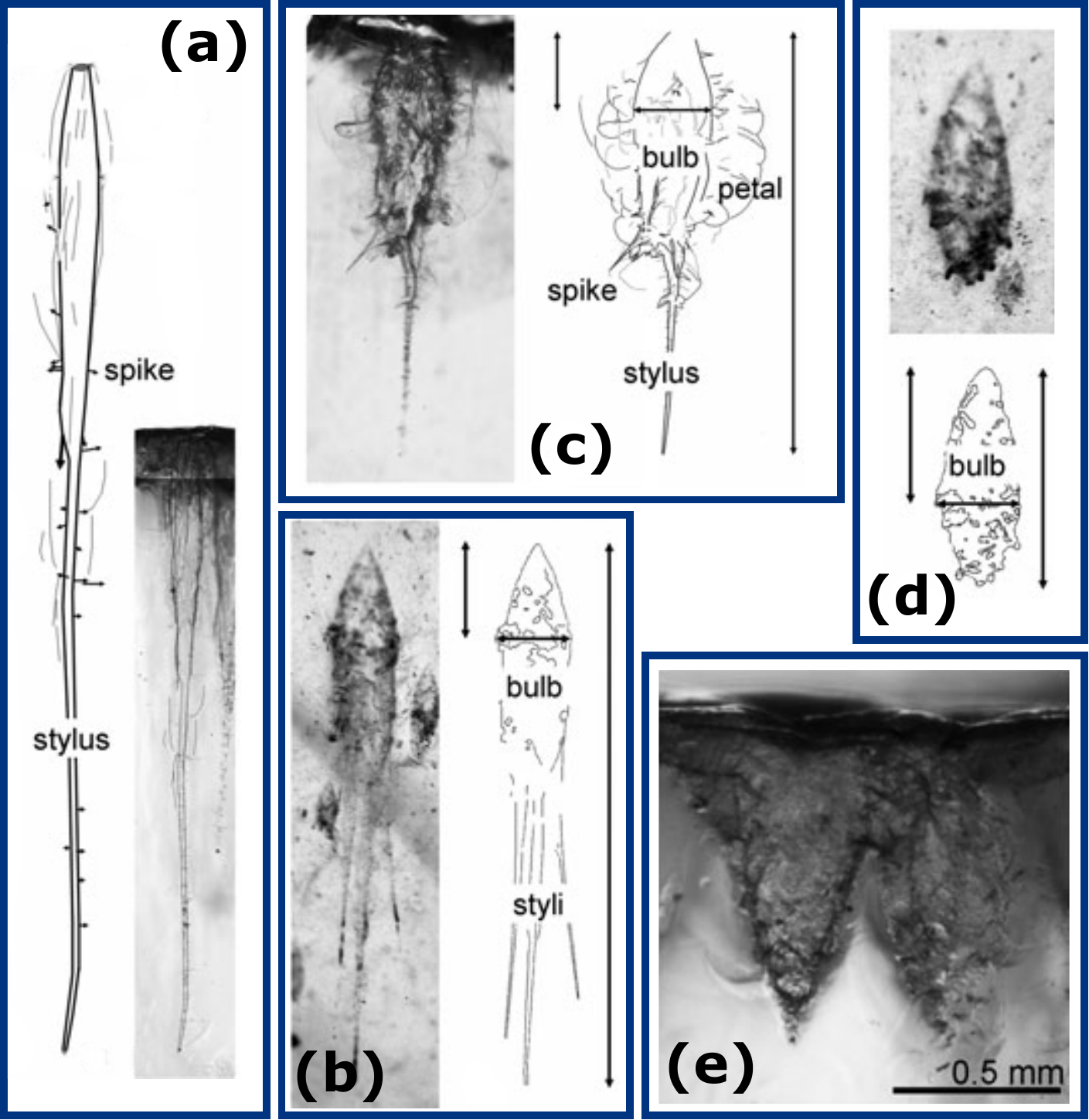}
  \caption{\label{fig:stardust_figure}Track morphology terminology for Stardust tracks, derived from \citet[][Fig. 1 and 12]{Kearsley_impact_2012}.
           (a) to (d) from aerogel tracks in Stardust collection, (e) from analogue experiment by \citet{Kearsley_impact_2012}.}
\end{figure}

Impacts into aerogel are divided into three main classes \citep{HoerzEtal:2006}:
Type A tracks are slender, flute- or carrot-shaped tapering tracks with either a single or multiple styli and/or spikes (cf. Fig. \ref{fig:stardust_figure} (a)), where a stylus is defined as that part of the particle track that runs about straight, looking like a narrowing tube or a root.
The shortest type A tracks, less than 100 \mum, were initially all classified as type A although it was then already noted that their morphology of a squat turnip is slightly different \citep{HoerzEtal:2006}.
Subsequent laboratory work with analogue material showed that their impactors are substantially different from the longer type A tracks, thus \citet{Kearsley_impact_2012} suggested to re-classify these tracks as type A* (cf. Fig. \ref{fig:stardust_figure} (e)).
Type B tracks show broader, bulbous cavities with one or several styli (cf. Fig. \ref{fig:stardust_figure} (b) and (c)), and type C tracks are broad, stubby cavities with no or very little styli (cf. Fig. \ref{fig:stardust_figure} (d)).

To determine impactor properties from track properties, many efforts of laboratory calibration were carried out, e.g. by \citet{Kearsley_impact_2012}.
Impactors that are suggested to match particles of the \SolidGroup\ in our classification are single crystals or glassy grains of sizes between 1 and 10 \mum\ \citep[note that the impactor particle diameters less than 1 \mum\ are all correlated to type A* tracks;][]{BurchellEtal:2008}.
These materials are not expected to fragment upon high-velocity capture and indeed are found to produce type A tracks with one stylus \citep{Kearsley_impact_2012}. 
However, type A tracks with one stylus can also be caused by agglomerates of up to 100 \mum\ with coarse subunits larger than about 10 \mum.
In our classification, this agglomerate would fall in the \PorousGroup.

The \PorousGroup\ is suggested to be populated by all agglomerate impactors used in the experimental calibrations, except for the most fragile ones \citep{Kearsley_impact_2012}.
These impactors of sizes between 1 and 100 \mum\ \citep{BurchellEtal:2008} were found to create type A tracks with single or multiple styli as well as type B tracks.
This ambiguity is in good agreement with the continuity between the track shapes of type A and B \citep{Kearsley_impact_2012} and possibly an effect of different aggregate strength depending on the degree of subunit fineness and organic content.

Extremely weak material with highest porosities as suggested in the \FluffyGroup\ was not used for laboratory calibrations and thus a comparison to the impacts created is difficult.
The best matches among the used calibration material is expected to be the agglomerates of fine subunits with organic material, or pure organic material.
Small impactors of few 100 nm in size were probably creating type A* tracks and those around 10 \mum\ \citep{BurchellEtal:2008} type C tracks. 

In comparison to the Rosetta/GIADA data, those impactors could well be the same materials as the particles creating fluffy detections without a sign of a compact particle in the GIADA instrument.
On the other hand, fluffy particles associated with the detection of a compact particle in GIADA (explained by \FluffySolidMix\ in Fig. \ref{fig:pictograms2}) are suggested to have caused Stardust type B tracks, where the fluffy part would create the bulbous morphology and the compact particle creates the stylus.

\subsubsection{Aluminium Foils}
\label{sect:StardustAlu}

Cometary dust particles that collided with the aluminium frame holding the aerogel collector produced hyper velocity craters and left molten residues inside \citep{HoerzEtal:2006}.
It was found in laboratory experiments that it is possible to deduce impactor properties such as size, mass, density and internal structure from crater morphology.

If the morphology of the craters is smoothly bowl-shaped, their suggested impactors are dense, 10 -- 60 \mum\ in size, thus could resemble the particles causing type A tracks in aerogel.
The residuals in the craters indicate that these particles must not be homogeneous in composition, but can also have consisted of a compact, about 3000 \density\ silicate particle accompanied by a fine grained material mix \citep{KearsleyEtal:2008}.
Thus, impactors creating bowl-shaped craters would fall into our \SolidGroup\ or \PorousGroup, or a mix of the two (e.g., \PorousSolidRimParticle\ or \PorousSolidFebo).

For craters with high and uneven relief \citet{KearsleyEtal:2008} suggest that they are caused by porous agglomerates with low densities, complex shape and diverse composition.
Their model calculations reveal porosities around 75 \% and densities less than 1000 \density, which classifies them as members of our \PorousGroup.
Agglomerate sizes can be up to 100 \mum, but their constituents are in the micrometre scale and seem to consist of again smaller particles in the tens of nanometres size range \citep{KearsleyEtal:2008}.
These smallest, tens of nanometres grains are falling into our \SolidGroup.

As there is no experimental data on extremely low-density and high-porosity material shot on aluminium foils \citep{KearsleyEtal:2008}, there is no counterpart for the \FluffyGroup\ in these laboratory studies.
Interestingly, the distribution of aerogel tracks and aluminium foil craters are only slightly consistent with random impacts and can be interpreted as clustering.
It was suggested that particles fragment in the coma, leading to so-called bursts and swarms in dust flux measurements \citep{TuzzolinoEtAl:2004, EconomouEtAl:2012}.
Clustering of impact features, be it aerogel tracks or aluminium foil impacts, could be the result of particle fragmentation, but the reason for breakup is unknown \citep{HoerzEtal:2006}.
If millimetre sized fluffy particles like the ones detected by Rosetta/GIADA (Sect. \ref{sect:GIADA}) were present at comet Wild 2, the aluminium foil clusters could even be explained without particle fragmentation, just by direct impact of fluffy particles or, if breakup is desired, the fragmentation of fluffy particles can be explained by electrostatic charging.

\subsection{Interplanetary Dust Particles and Micrometeorites}
\label{sect:IDP}

The largest sample of cometary material on Earth is believed to be in Interplanetary Dust Particles (IDPs) and Micrometeorites (MMs).
While IDPs are particles collected in the Earth stratosphere, MMs are collected on ground (e.g., Antarctica, sediments, ...).
The association with cometary material is not unambiguous but several arguments support it on a statistical level.

The zodiacal cloud model of \citet{NesvornyEtal:2010}, suggests from a dynamical perspective that 85 \% of the total mass influx at Earth originates from Jupiter family comets.
Particles smaller than $\sim 300$ \mum\ should moreover survive frictional heating to arrive in the Earth stratosphere (as IDPs) and even on the surface (as MMs).
Interpretations of zodiacal light observations in the visible and infra-red domains \citep{LasueEtal:2009, RowanRobinsonMay:2013} also indicate that most of the interplanetary dust particles reaching the Earth's vicinity are of cometary origin.

\citet{BusemannEtal:2009} connected IDPs collected in April 2003 with comet 26P/Grigg-Skjellerup, which was expected to show an enhanced flux in this time period.
From a compositional standpoint, this sample shows very primitive properties:
an unusually high abundance of pre-solar grains, organic matter, and fine-grained carbonates.
As a classification, \citet{Bradley:2003} and \citet{Rietmeijer:2002} distinguish IDPs between \emph{chondritic} and \emph{non-chondritic} material.
\citeauthor{Bradley:2003} further differentiates the morphological appearance into \emph{chondritic porous} (CP) and \emph{chondritic smooth} (CS).
The chondritic porous particles are considered the most primitive IDPs, being composed of unhydrated phases and not showing products of aqueous alteration.
The IDPs described by \citet{BusemannEtal:2009} fall into this chondritic porous group.

The measurement of physical properties -- we are particularly interested in porosity and strength -- are not easy on this scale.
To our knowledge, nothing is published on the strength but there were some attempts to measure porosities \citep[also see summary in][]{Rietmeijer:1998}.
One possibility of IDP porosity measurement is described by \citet{LoveEtal:1994}:
The volume was determined with a combination of transmission and scanning electron microscopy (TEM and SEM).
For the mass determination, they measured the Fe count rate of IDPs from x-ray fluorescence \citep[see also][]{FlynnSutton:1991} and enhance this mass with the assumption of a chondritic composition.
They confirm with calibration measurements that in their measurement range of $5 - 15$ \mum\ the entire volume is excited.
The determined density distribution is corrected for a size bias in fall speed (coupling times).
The corrected numbers show a large proportion of particles with densities near and below 2000 \density\ (40 \% porosity) but also higher densities (up to 6000 \density) were observed.

For larger samples of chondritic IDPs in the 10 -- 100 \mum\ range, \citet{CorriganEtal:1997} produced thin sections and measured direct porosity from SEM images.
Their results peak around 4 \% and the tail of the distribution (though small numbers) range to 53 \% porosity.


It seems as if our \emph{solid} and \PorousGroup\ from Sect. \ref{sect:StructurePorosityClassification} are represented in the IDP collections.
Moreover, there are indications for higher porosities \citep{Rietmeijer:1993}, indicating members of the \FluffyGroup.

\begin{figure}[t]
  \centering
  \includegraphics[width=\columnwidth]{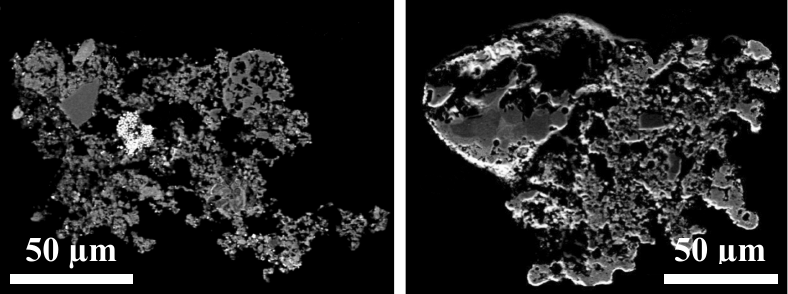}
  \caption{\label{fig:MMs}SEM image of micrometeorites (MMs) by \citet{GengeEtal:2008}.
           Left: Their fine grained C3 type, associated to our \PorousGroup.
           Right: Particle consisting of coarse grained (top left) and fine grained (bottom right) components, similar to mixes we have exemplified in Fig. \ref{fig:pictograms2}, in particular \PorousSolidFebo.}
\end{figure}

To further link these studies to our morphologic classification, we look into the classification of MMs.
Their link to cometary material is weaker than the IDPs' but plenty of material exists and a classification was presented by \citet{GengeEtal:2008}.
In their Table 1, they define three groups from the MMs' melting state, where we are mostly interested in their \emph{unmelted MMs}.
Further they define the sub-classes \emph{fine grained}, \emph{coarse grained}, \emph{refractory}, and \emph{ultracarbonaceous}.
The fine grained MMs can be reasonably porous (C3 in their nomenclature, example in Fig. \ref{fig:MMs}, left), falling into our \PorousGroup.
Their coarse grained MMs would fall into our \SolidGroup\ but due to possible alteration of the MMs we should be careful in connecting these with cometary dust.
Also \citet{GengeEtal:2008} found that particles might not fall into a unique class as their example from Fig. \ref{fig:MMs} (right) shows.
This mix of coarse grained and fine grained material resembles our examples of mixed cases in Fig. \ref{fig:pictograms2}, in particular \PorousSolidFebo.
The refractory and ultracarbonaceous MMs classified by \citeauthor{GengeEtal:2008} are likely associated with cometary dust but are classified by compositional arguments why we do not draw a comparison here.

\subsection{Earth-based Observations of 67P and other Comets}
\label{sect:EarthObs}

\subsubsection{Dust Tail and Trail}

Observation and modelling of the dust tail and trail can provide independent clues on particle sizes present in the coma.
The shape of tail and trail from Earth-based observations is mostly determined by (a) the size- and time dependent dust production rate and velocity and (b) forces acting on these dust particles \citep[for a review on this topic see][]{AgarwalEtal:2007}.
The first is a free parameter constrained by Rosetta, the for the latter, the decisive quantity is the ratio of solar radiation pressure and gravity that depends on the size, optical properties, and bulk density of a particle \citep{BurnsEtal:1979}.
Tail models are more sensitive to small particle sizes while the trail is dominated by large particles ($\gtrsim 100$ \mum).
\citet{MorenoEtal:2017} used 116 tail observations from various telescope plus trail observations from the last and previous apparitions of comet 67P.
With a Monte Carlo model, they propagated the trajectories of dust particles and compared thus generated synthetic images to the available observations.
Motivated by the small amount of (sub-)micrometre particles detected by MIDAS, they used 20 \mum\ as a minimum particle diameter.
They let the maximum particle diameter vary from 2 to 80 cm, as a function of heliocentric distance (largest size around perihelion).
Also the slope for particles with < 2 mm diameter varies with heliocentric distance.
No further evolution of particle sizes in the coma such as fragmentation or sublimation is applied nor required.
Overall, they reached a good fit to the telescope imaging data and also the dust production rate -- an outcome of the model -- shows a reasonable correlation, albeit not complete, with gas production rates \citep{HansenEtal:2016}.

The size input parameters, albedo, and bulk density are chosen to be consistent with Rosetta.
However, the fit to the telescope data confirms their applicability also on the large scale.
In particular the minimum required particle size of 20 \mum\ is interesting.
Smaller particles do exist, i.e., they can be dispersed and were observed \citep[and others]{Bockelee-MorvanEtal:2017, Bockelee-MorvanEtal:2017erratum}, but they are too few to manifest themselves in the scattered light of telescope observations.
In particular, the steep size distribution adopted for small particles near perihelion (differential size distribution exponent $\leq -3.5$) implies that the scattering cross-section is concentrated in the smallest particles near the minimum adopted size.
This implies that the cross section of particles $<20$ \mum\ must be small (i.e., a sharp cut-off in the size distribution around this size), as otherwise either the model coma would become much brighter than the observed coma, or the model dust production rate of larger particles would have to be decreased to become inconsistent with Rosetta observations and the surface brightness of the trail.
This lack of $<20$ \mum-diameter particles is in line with earlier findings at 67P \citep{AgarwalEtal:2010, FulleEtal:2010} and with Spitzer observations of a larger sample of Jupiter Family Comets \citep{ReachEtal:2007}.
An assessment of the minimum required and maximum allowed population for these sizes would be interesting but is currently not available.

\subsubsection{Polarimetry}

Remote observations of solar light scattered by dust in cometary comae and tails have been used for more than a hundred years to study its partial linear polarisation, as first done on comet Tralles in 1819 \citep{Arago:1858}.
The linear polarization is connected to chemical composition and physical properties of the dust.
During 1P/Halley's return in 1985-1986, the evolution of the polarisation pointed towards significant changes in dust properties during outbursts \citep[Earth-based;][]{DollfusEtal:1988} as well as variations related to jet structures \citep[when Giotto crossed these;][]{LevasseurRegourdEtal:1999, FulleEtal:2000}.
Polarimetric remote observations of comet 67P have been performed during its 2008-2009 and 2013-2016 returns in preparation and support of the Rosetta mission \citep{HadamcikEtal:2010, HadamcikEtal:2016, RosenbushEtal:2017}.
The interpretation of such datasets requires experimental and numerical simulations to infer information on chemical composition and physical properties of the dust observed in cometary comae \citep[e.g.,][]{LevasseurRegourdEtal:2007}.
Experimental simulations on numerous samples are obtained with gonio-polarimeters, operating in the laboratory and/or under micro-gravity conditions \citep{MunozHovenier:2011, MunozHovenier:2015, LevasseurRegourdEtal:2015}.

Agglomerates of sub-micrometre sized grains best fit the higher polarisation observed in cometary jets and after fragmentation or disruption events, while a mixture of porous agglomerates and compact particles are needed to fit whole comae observations \citep{HadamcikEtal:2006}.
The polarimetric phase curves of cometary analogues made of porous agglomerates of sub-micron-sized Mg-silicates, Fe-silicates and carbon black grains mixed with compact Mg-silicates grains are comparable to those observed in comae of comets \citep{HadamcikEtal:2007}.
Numerical simulations complement the experimental work to infer further dust properties:
It has been established that spheres or spheroids, in fact any solid particles, cannot (even with various size distributions and compositions) reproduce the observational data \citep{KolokolovaEtal:2004} and that models with agglomerates of sub-micrometre sized particles provide satisfactory results \citep{KiselevEtal:2015}.
Simulations indeed strongly suggest that cometary dust is a mixture of (possibly fractal) agglomerates and of compact particles of both non-absorbing silicates-type materials and more absorbing organic-type materials \citep{Levasseur-RegourdEtal:2008, LasueEtal:2009}.

The variety of agglomerate and grain structures is thus consistent with the scheme developed in Sect. \ref{sect:StructurePorosityClassification}, with an emphasis on the porous group.

\section{Discussion and Interpretation}
\label{sect:discussion}

Using the morphologically motivated classification scheme from Sect. \ref{sect:StructurePorosityClassification}, we have in Sect. \ref{sect:InstrumentResults} summarised recent results on cometary dust, which culminated in Table \ref{tab:table}.
We now want to change the view point and discuss the results in a comparative manner and exploit the potential from the complementary design of the individual instruments.

The imaging capabilities of MIDAS and COSIMA provide similar insights on a different but overlapping size scale.
Both show the surface -- and to some level interior -- structure of porous dust agglomerates (\PorousGroup\ in Fig. \ref{fig:pictograms}), possibly with sub-structure as in \PorousAgglomerateCluster.
The MIDAS capability of strength measurement is not yet fully exploited but from COSIMA we learned that all agglomerates break when applying the SIMS ion beam.
In contrast to that, the measurement parameters of GIADA are size (relying on a light scattering model) and mass, from which we can derive densities and porosities.
Sizes are also determined by OSIRIS and VIRTIS:
While OSIRIS provided insight on large dust particles up to a metre (again, using a light scattering model), VIRTIS provided constraints on the power law size distribution and smallest unit size from modelling.

\begin{figure}[t]
  \centering
  \includegraphics[width=\columnwidth]{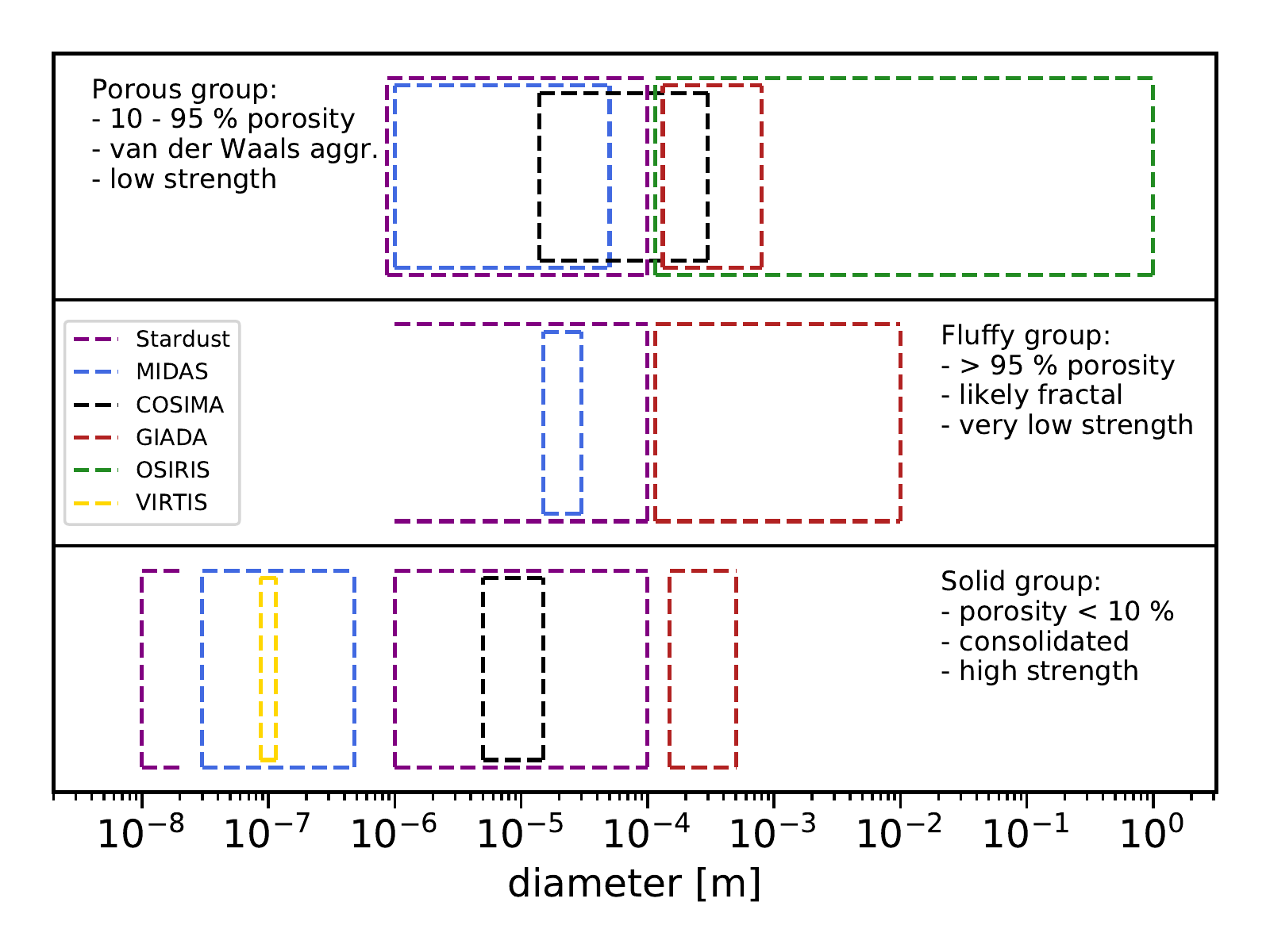}
  \caption{\label{fig:OverviewPlot}Visual representation of Table \ref{tab:table} to show where different instruments have overlaps in their sensitivity range.
	         Open boxes denote unknown size limits, i.e., smaller equal or larger equal than the plotted size.}
\end{figure}

In Fig. \ref{fig:OverviewPlot} we provide a visual representation to allow for a more quantitative comparison of the observed size ranges.
Every colour represents an instrument (as in Table \ref{tab:table} we include Stardust in the comparison) and the results are grouped by the three groups defined in Sect. \ref{sect:StructurePorosityClassification}.
A lot of overlap is seen for porous agglomerates in the COSIMA size range (14 -- 300 \mum, black dashed box on the top), which overlaps with MIDAS, GIADA, OSIRIS, and Stardust.
The overlap confirms that we have the best complementary knowledge for aggregates in this size range.
However, it has to be noted that measurements and interpretations (e.g., porosities) are not overall consistent in spite of falling into the same group.
Given that instruments and also interpretations can have biases, this overlap has potential to further understand and correct for these.

In the same box (\GroupPorous), there are no agglomerates smaller than 1 \mum.
This is a choice we made in Sect. \ref{sect:MIDAS}, where we interpreted sub-micrometre structures scanned by MIDAS as solid components, which is possible but ambiguous.
The size of the smallest solid component is an interesting topic with important implications for solar system formation.
One particularly noteworthy contribution here is the detection of 100 nm sized particles needed to explain the observation of super heating in VIRTIS spectra (Sect. \ref{sect:VIRTIS}).
We do have imagery evidence of structures on this size scale from MIDAS (Fig. \ref{fig:MidasSmallest}):
The cauliflower structure in Fig. \ref{fig:MidasSmallest} is interpreted as a porous agglomerate ($\sim 1$ \mum) of highly irregular but solid aggregates (100 nm size range).
It becomes evident that we are lacking data on the $\lesssim 1$ \mum\ size range and we are left with indirect evidence, which should be the focus of upcoming research.

Large particles from the \SolidGroup\ were described by Stardust (aerogel and aluminium foil; 1 -- 100 \mum), COSIMA (CAI candidate and specular reflection; 5 -- 15 \mum) and GIADA (measurements of high densities; 0.15 -- 0.5 mm).
There is no strong evidence for solid particles larger than 1 mm but there is also no robust method to discriminate between porous agglomerates and solid particles in the OSIRIS size range.
Our choice of group is supported by indirect density measurements on the (sub-)millimetre scale \citep{GuettlerEtal:2017, FulleEtal:2018} but a small fraction of solids could have been unrecognised.

Agglomerates from the \FluffyGroup\ smaller than 1 cm are found by MIDAS and GIADA, confirming Stardust findings.
Their existence as a small fraction of the full dust population is plausible within the context of planetesimal formation \citep{FulleBlum:2017}.
In fact the complete picture as presented in Fig. \ref{fig:OverviewPlot} can be consistent with primordial dust agglomeration:
In the comet forming region, we would expect small solid particles, either falling in from the interstellar cloud or transported out from the inner solar system.
These coagulate into fractal agglomerates if allowed by the dynamic environment.
Larger agglomerates would restructure in collisions while they grow, such that we expect everything above a threshold size (e.g., mm size range, depending on model) to fall into the \PorousGroup.

\begin{figure}[t]
  \centering
  \includegraphics[width=\columnwidth]{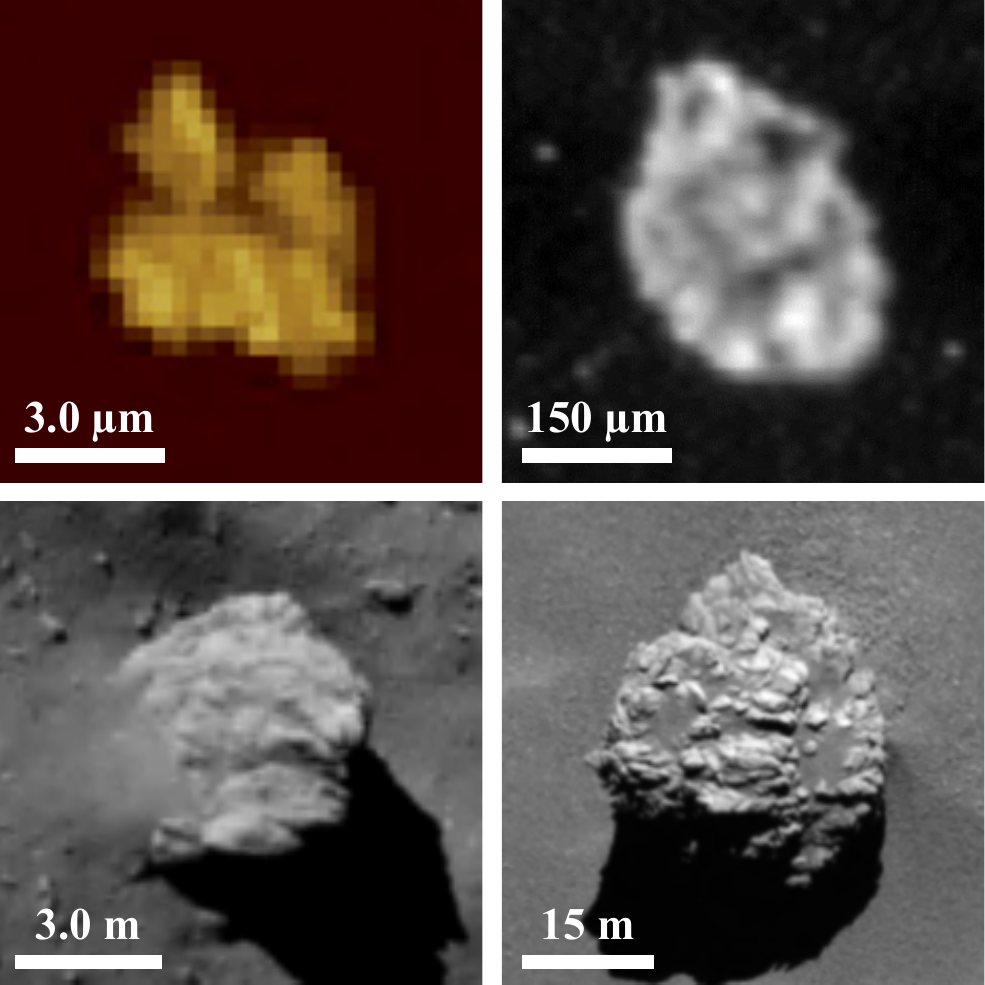}
  \caption{\label{fig:SelfSimilarity}Agglomerates and boulders of comet 67P at different scales from MIDAS \citep[top left,][]{bentley_midas_2016}, COSIMA \citep[top right,][]{LangevinEtal:2016}, ROLIS \citep[bottom left,][]{MottolaEtal:2015}, and OSIRIS (bottom right, NAC image at \mbox{2015-02-14T15:31:05}).}
\end{figure}

Another aspect linked to the primordial growth of these agglomerates is their structure and porosity.
Not only does the structure observed by MIDAS and COSIMA show similarities in the build-up from smaller sub-structures.
The same in principle applies to surface boulders observed by OSIRIS and ROLIS.
Figure \ref{fig:SelfSimilarity} shows the comparison of MIDAS and COSIMA agglomerates (top) and ROLIS and OSIRIS boulders (bottom).
It is interesting that morphologies on these different size scales -- from sub-micrometre to tens of metres, i.e., over 7 orders of magnitude -- appear similar.
At least the organisation into sub-structures is evident on all scales.
It is tempting to apply the concept of \PorousAgglomerateCluster\ in Fig. \ref{fig:pictograms} and extend this to form clusters out of clusters and so forth in a hierarchic structure.
The problem with this is that one would expect to add porosity at every step of assembly:
If the initial agglomerates (\PorousAgglomerate) have a filling factor (1 -- porosity) of 0.5, the first cluster of agglomerates (\PorousAgglomerateCluster) has a filling factor of $0.5^2$, a cluster of these has a filling factor of $0.5^3$ and so forth.
Assuming a size ratio of 10 between cluster sizes and their next smaller component, the filling factor of a 10 m boulder would be $0.5^7 = 0.008$ (99.2 \% porosity).
This is of course absurd and widely inconsistent with the comet's bulk porosity \citep{SierksEtal:2015}.
This exemplifies that structure and porosity have to be treated as individual parameters.
It is possible that the structure is the result of an agglomeration process that also involved compaction.
In that case, the porosity would be found on the smallest and strongest scales.
Porosity as a function of size would thus increase for small sizes until it remains constant at a threshold size, which is expected to be close to the comet's bulk porosity.

\section{Conclusions}
\label{sect:conclusion}

This article presented the first summary and inter-comparison of results from all Rosetta dust instruments.
We established a classification scheme (Sect. \ref{sect:StructurePorosityClassification}) based on structure, porosity, and strength of fluffy and porous dust agglomerates, compact aggregates, and grains.
This classification was compared to results of Stardust and also Earth observations of probable cometary dust.
These include tail and trail observations as well as polarimetric studies.
Also the information of IDPs and MMs was reviewed from the standpoint of our classification.

Given that different instruments and methodologies have different measurement parameters, resulting in different descriptions of their results, it is a success to be able to describe this amount of data, all within the same framework of our classification.
The choice to constrain our classification to morphology, porosity, and strength is a large restriction and can only be a first step in a complementary study of cometary dust after Rosetta.
The work needs to be continued and extended and we hope that the presented classification will help.

\section*{Acknowledgement}
We thank all Rosetta instrument teams, the Rosetta Science Ground Segment at ESAC, the Rosetta Mission Operations Centre at ESOC and the Rosetta Project at ESTEC for their outstanding work enabling the science return of the Rosetta Mission.

\bibliographystyle{aa}
\bibliography{literature}

~\\{
  \small
  Compiled on \today\ from
  \IfFileExists{\gitfilename}{GIT (\gitrevision).}{overleaf.}
}

\end{document}